\documentclass[twocolumn,preprintnumbers,amsmath,amssymb]{revtex4-2}

\usepackage{graphicx}
\usepackage{dcolumn}
\usepackage{bm}
\usepackage{siunitx}
\usepackage{xcolor}
\usepackage{placeins}
\usepackage[normalem]{ulem}
\usepackage[colorlinks]{hyperref}
\usepackage{mathrsfs}
\usepackage{blindtext}

\begin{document}

\title{Neutralization of impurities by cw-excitation of high-$n$ Rydberg-excitons}
\author{J. Heck\"otter$^1$, B. Panda$^1$, K. Br\"agelmann$^1$,   and M. A\ss mann$^{1}$}
\affiliation{$^{1}$ Experimentelle Physik 2,
                Technische Universit\"at Dortmund,
                D-44221 Dortmund, Germany}

\date{\today}

\begin{abstract}
We investigate the neutralization of charged defects in a Cu$_2$O crystal experimentally with a cw two-color pump-probe setup. We demonstrate that the excitation of high-lying Rydberg excitons at low densities generates an enhancement of exciton absorption up to a factor of 3. The result is an overall improvement of the absorption spectrum's quality in terms of increased oscillator strengths and narrower linewidths, referred to as \textit{purification}. 
The reduction of the amount of charged impurities in the crystal is further verified by a thorough analysis of impurity photoluminescence signals. 
\end{abstract}

\maketitle

\section{Introduction}
Rydberg excitons are highly-excited pairs of electrons and holes in a semiconductor. They can reach extensions in the $\mu$m range and, therefore, are mesoscopic quantum objects. They represent the analogues of Rydberg atoms embedded in a solid-state environment and attracted huge interest in a variety of research fields, since their discovery in 2014~\cite{kazimierczukGiantRydbergExcitons2014c} in the material Cu$_2$O. 
Starting from fundamental exciton physics in external fields~\cite{schweinerMagnetoexcitonsCuprousOxide2017a, rommelMagnetoStarkEffectYellow2018, zielinska-raczynskaElectroopticalPropertiesRydberg2016, zielinska-raczynskaMagnetoopticalPropertiesRydberg2017, kurzexcitonic2017} and quantum chaos~\cite{assmannQuantumChaosBreaking2016, schweinerMagnetoexcitonsBreakAntiunitary2017}, Rydberg excitons have also led to new discoveries in the fields of long-range interactions~\cite{waltherInteractionsRydbergExcitons2018a,heckotterAsymmetricRydbergBlockade2021} and quantum many-body phenomena~\cite{semkatInfluenceElectronholePlasma2019, semkatQuantumManybodyEffects2021, waltherPlasmaEnhancedInteractionOptical2020} as well as strong optical non-linearities~\cite{morin_self-kerr_2022, zielinska-raczynskaNonlinearOpticalProperties2019,  zielinska-raczynskaElectromagneticallyInducedTransparency2016,waltherElectromagneticallyInducedTransparency2020} and Rydberg-polaritons in cavities~\cite{waltherGiantOpticalNonlinearities2018a,orfanakisRydbergExcitonPolaritons2022}.

In  Cu$_2$O, the excitonic transitions between the highest valence and the lowest conduction band are called the yellow series, referring to the excitation wavelength range between 610 and 570~nm. Since both bands are of positive parity, direct transitions of even parity $S$ and $D$ excitons are dipole-forbidden, but weakly allowed for odd-envelope $P$ excitons, resulting in small oscillator strengths and  narrow linewidths. Together with a comparably high Rydberg energy of 90~meV, these properties render it possible to observe an extended excitonic Rydberg series up to principal quantum numbers above $n=20$ in natural high-quality crystals~\cite{kazimierczukGiantRydbergExcitons2014c, heckotterExperimentalLimitationExtending2020}. 
Recently, even the observation of Rydberg excitons with quantum numbers as high as $n=30$ was reported in a photoluminescence excitation scheme~\cite{versteeghGiantRydbergExcitons2021}. Excitons in these highly-excited states reach extensions in the $\mu$m range.

Besides strain and temperature, one of the parameters that determine the highest observable Rydberg state  in a sample is the density of charged impurities in the crystal, that varies slightly from sample to sample~\cite{lynchRydbergExcitonsSynthetic2021}.  Moreover,  charged impurities cause a broadening of linewidths and a reduction of oscillator strengths, in particular for the highest $n$, as has been shown theoretically in Ref.~\cite{krugerInteractionChargedImpurities2020}. 
Due to their large polarizability that scales with the principle quantum number as  $n^7$~\cite{heckotterScalingLawsRydberg2017a}, Rydberg excitons are highly sensitive to the presence of small electric fields stemming from charged defects with densities on the order of $0.001$~$\mu$m$^{-3}$~\cite{krugerInteractionChargedImpurities2020}. 
Consequently, the excitonic line shape parameters measured in Refs.~\cite{kazimierczukGiantRydbergExcitons2014c, heckotterExperimentalLimitationExtending2020, versteeghGiantRydbergExcitons2021} do not follow the theoretical predictions in the high $n$ regime. In particular, both oscillator strengths and linewidths are expected to scale with the principle quantum number as $n^{-3}$~\cite{elliottIntensityOpticalAbsorption1957, stolzInteractionRydbergExcitons2018}. However, the measured oscillator strengths are found to drop more strongly with $n$, while the linewidths are found to be inhomogeneously broadened, which renders experimental access to high-$n$ states challenging.  

Here, we show an experimental approach to overcome this issue of line parameters not reaching ideal scalings by neutralizing a fraction of the charged impurities in a sample. 

In other semiconductor quantum technology platforms, like semiconductor quantum dots, charge fluctuations are known to have a strong influence on the coherence properties of optical transitions and to cause spectral shifts or inhomogeneous broadening of emission lines~\cite{reigueResonanceFluorescenceSingle2019}. 
A well-established technique to circumvent this problem is  additional non-resonant laser excitation, typically tuned to an energy above the band gap, effectively stabilizing the electrostatic environment around a quantum dot~\cite{nguyenOpticallyGatedResonant2012, houelProbingSingleChargeFluctuations2012, haucklocating2014, chenCharacterizationLocalCharge2016, makhoninwaveguide2014}. 
However, in Cu$_2$O, the electron-hole plasma created by additional above-band gap excitation is known to cause a band gap shift leading to the disappearance of Rydberg exciton lines even at ultra-low densities of around 0.01~$\mu$m$^{-3}$~\cite{heckotterRydbergExcitonsPresence2018}, but not to improve the spectral quality. 

We show that weak additional cw-laser excitation at a particular energy within a narrow spectral range around the band gap creates a low density of high-$n$ Rydberg excitons that interact with charged impurities, eventually neutralizing them, which enhances the oscillator strengths and brings the linewidths closer to the Fourier-transform limit. 
This effect is first reported in Ref.~\cite{bergenTamingChargedDefects2023} with a focus on the dynamics studied by time-resolved measurements of the relative transmission. The present manuscript focuses on cw absorption spectra providing access to high spectral resolution and to absolute changes in the absorption coefficient. 

The paper is organized as follows: In Sec.~\ref{experiment} the experimental setup is described. Sec.~\ref{results} forms the main part of this manuscript.  The section starts with a general description of the observed enhancement of absorption in~\ref{enhancement}. The following subsections discuss the enhancement as a function of excitation power~\ref{power} and as a function of excitation energy~\ref{wavelength}. In the last subsection, Sec.~\ref{sample2}, a sample of lower quality is studied. Section~\ref{PL} shows the change of photoluminescence signals of impurities induced by an additional pump laser. Finally, conclusion and an outlook are given in Secs.~\ref{discussion} and \ref{outlook}. 

\section{Experiment}\label{experiment}
\begin{figure}
	\begin{center}
		\includegraphics[width=\columnwidth, draft=false]{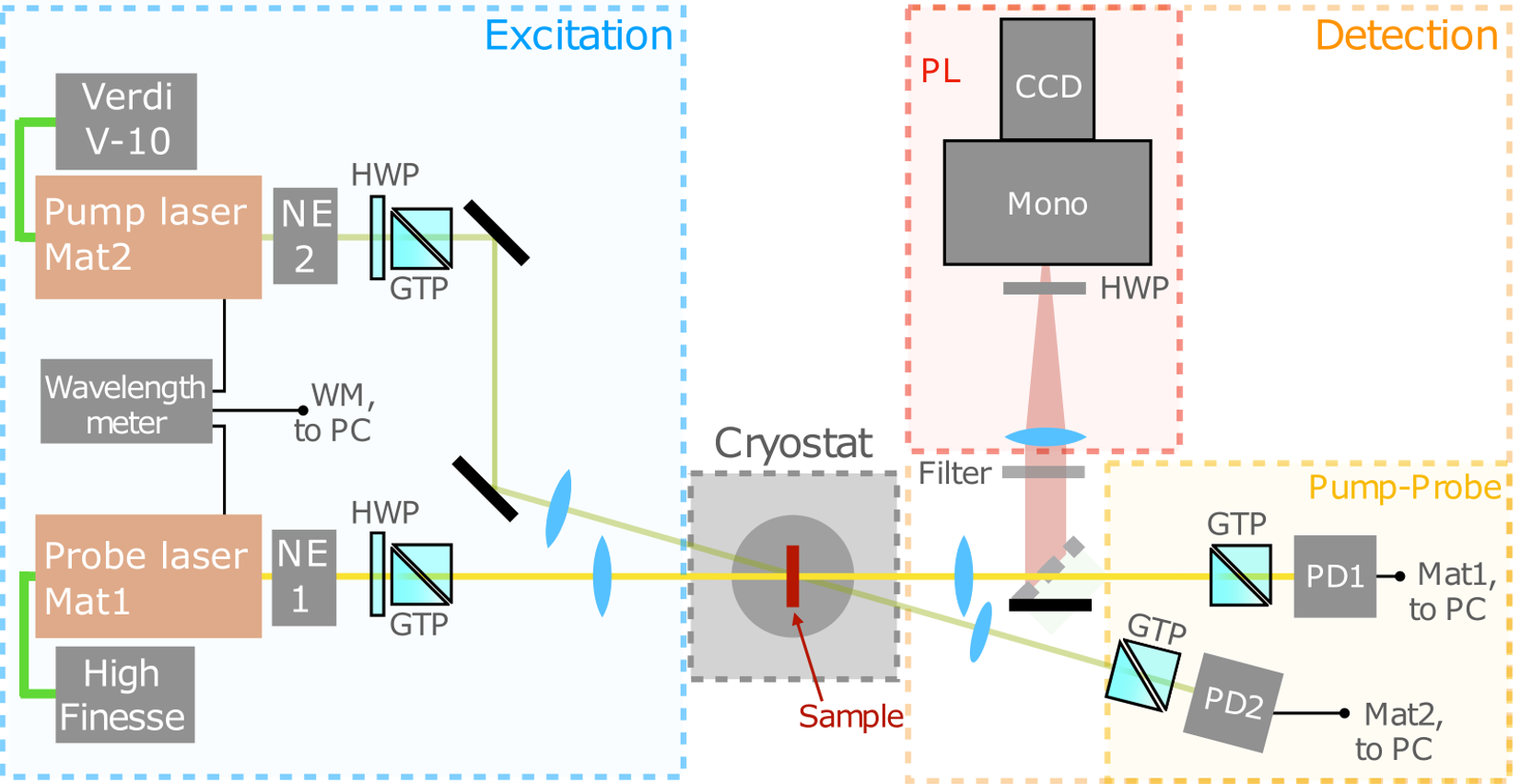} 
	\end{center}
	\caption{Scheme of the experimental setup: Two tunable dye laser are used to excite the sample. Both lasers beams are detected by photodiodes. Photoluminescence is detected by a combination of monochromator and CCD camera. Abbreviations: WM: wavelength meter; NE: noise eater; HWP: half-wave plate; GTP: glan-taylor prism;  PD: photodiode; Mat1/2: Matisse 1, Matisse 2; CCD: charge-coupled device camera. 
	} 
	\label{fig:Exp}
\end{figure}

The experimental setup used is sketched in Fig.~\ref{fig:Exp}. It is based on the setups shown in Ref.~\cite{kazimierczukGiantRydbergExcitons2014c}, but extended in the detection pathway. 
For excitation,  two tunable continuous wave (cw) dye lasers (Sirah Matisse DS) are exploited in a pump probe setting. The intensity of both beams is stabilized by noise-eaters (Brockton LPC). Further, they are focused on the sample, with the pump spot being 3 times larger than the probe beam (300 $\mu$m and 100 $\mu$m) to ensure a homogeneously distributed  pump intensity across the probe spot.  Furthermore, the overlap of both beams is controlled and routinely monitored by maximization of the pump-induced change of the probe signal. Both lasers are linearly polarized with crossed polarizations in order to remove stray light from signal with Glan-Taylor prisms. 
For detection, two different approaches are used:
First, the  light  of both pump and probe beam transmitted through the sample is recorded with a photodiode (New Focus Large-Area Photoreciever 2031) each, which renders the simultaneous detection of both signals possible. 
Second, the photoluminescence (PL) signal from the sample in the spectral range from 900-600 nm is detected with a spectrometer (Acton SpectraPro-500i) providing a spectral resolution of 4~meV.  

The investigations reported here are based on two samples of Cu$_2$O. The majority of experiments is performed on sample I, a high-quality natural Cu$_2$O sample of 34~$\mu$m thickness, mined in Namibia, cf. Ref.~\cite{kazimierczukGiantRydbergExcitons2014c}. The second one, sample II, is of inferior quality, and has been artificially grown by Schwab et al.~\cite{loison_progress_1980, bloch_sample_1980}. It has a thickness of 53~$\mu$m. The results on sample II are discussed in section~\ref{sample2}. For all experiments, the samples are cooled in a cryostat to 1.3~K by pumping on a superfluid He$^4$ bath. 

\section{Results}\label{results}
\subsection{Enhancement of absorption}\label{enhancement}
\begin{figure*}
	\begin{center}
		\includegraphics[width=1\textwidth]{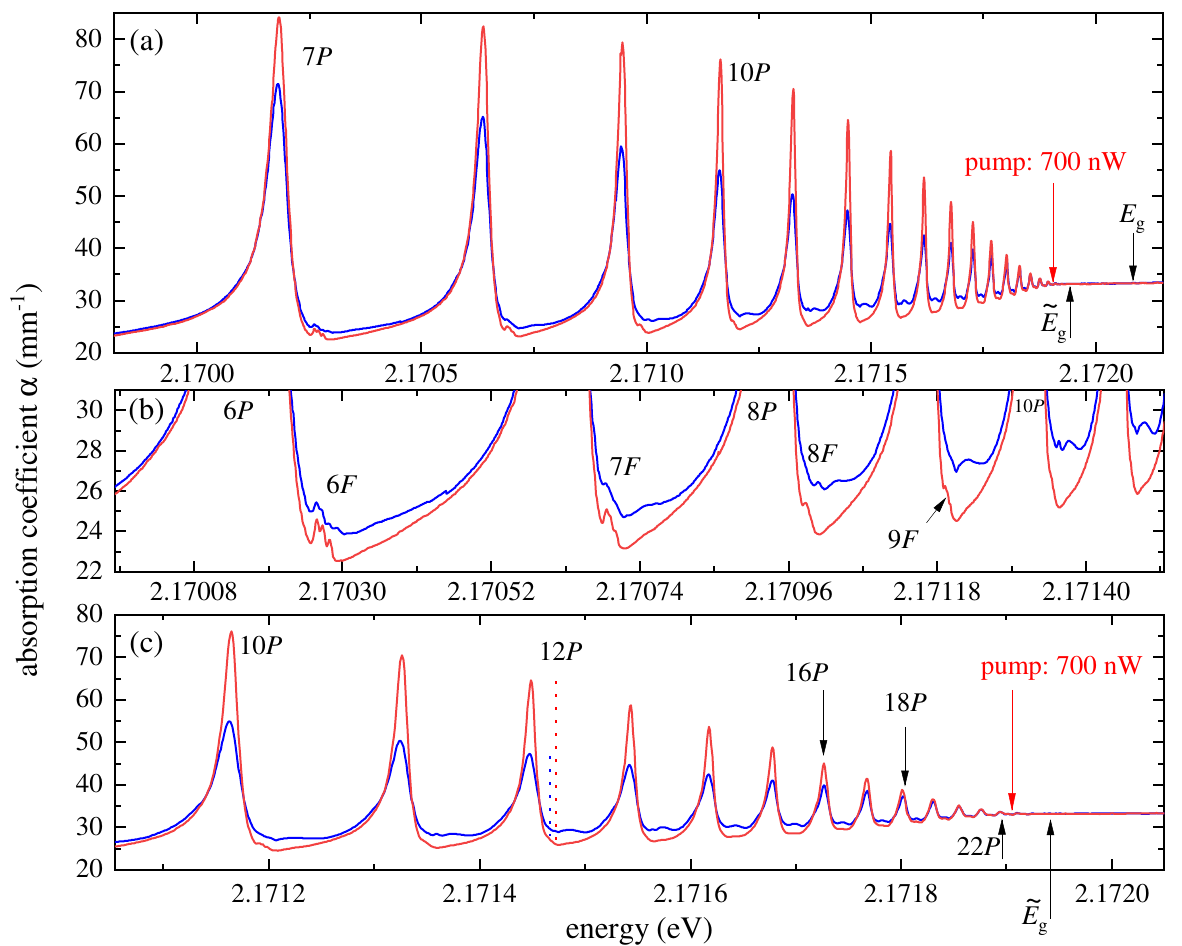} 
	\end{center}
	\caption{(a) Absorption spectrum of Rydberg excitons from $n=7$ to the band gap in the absence of a pump laser (blue) and with an additional illumination of 700 nW at an energy in the vicinity of the apparent band gap
		$\tilde{E}_\text{g}$ (red). The probe power is set to 200~nW. The pump laser induces a growth of oscillator strengths and a narrowing of lines resulting in an increased peak absorption. 
		(b) Focus on low-$n$ states $n=6$ to $n=10$: The absorption between adjacent $P$ excitons decreases due to the presence of the pump laser. Consequently, individual $nF$ absorption lines become more pronounced. 
		(c)  Focus on high-$n$ states: An increase of absorption is observed up to $n=18$, while for higher states no change can be observed. 
		The peak absorption of exciton lines from $n=11$ to $n=15$ increases by a factor of 2, indicated by the vertical red and blue dashed lines next to $n=12$. 
		The total number of visible $P$ exciton lines, $n_\text{max}$, does not change. }
	\label{fig:PerfectSpec}
\end{figure*}

First, we discuss absorption spectra where we scan the probe laser energy along the Rydberg series while the pump laser is fixed at a specific energy close to the band gap. 
The blue curve in Fig.~\ref{fig:PerfectSpec} shows a typical absorption spectrum of the high-quality sample I without additional pump laser illumination starting from the exciton resonance $n=7$ up to energies above the band gap. The spectrum is measured with a low probe power of 200~nW. The peak absorption at each individual exciton resonance decreases with increasing principal quantum number $n$ up to an $n_\text{max}$ which depends on the gemstone quality and also on the particular position on the studied sample~\cite{heckotterExperimentalLimitationExtending2020}.  
Here,  the highest observable exciton line corresponds to the principal quantum number $n_\text{max}=24$. 
On the high energy side of this state the spectrum transforms into a continuum showing a flat, weakly increasing absorption. 
This spectral position is typically called the apparent band gap  $\tilde{E}_\text{g}$~\cite{heckotterRydbergExcitonsPresence2018}, that is shifted by around 140~$\mu$eV below the nominal band gap $E_\text{g}=2.17208$. 
Here, $\tilde{E}_\text{g}=2.17194$~eV.  

In earlier publications, exciton-exciton~\cite{heckotterAsymmetricRydbergBlockade2021} or plasma-exciton~\cite{heckotterRydbergExcitonsPresence2018} interactions were studied in a similar pump-probe setting by tuning the pump laser to a specific wavelength or energy and measuring the change of probe laser absorption as a function of pump power. For these experiments, typical pump powers required to observe the effects of these interactions reached several $\mu$W up to tens of mW.
For the kind of interaction we study here, pump laser powers as low as 700~nW are sufficient to achieve the largest observable effects under comparable conditions, whereas the pump energy is put close to the apparent band gap $\tilde{E}_\text{g}$ as indicated by the red arrow. 
The resulting spectrum is shown by the red curve in Fig.~\ref{fig:PerfectSpec}(a). 
A comparison to the unpumped spectrum reveals the improved quality of the absorption spectrum: For all excitonic lines the absorption increases in presence of the pump laser beam. Also, the absorption in between adjacent lines decreases. As a consequence, the $F$ exciton triplets, typically visible from $n=4$ to $n=7$~\cite{thewesObservationHighAngular2015}, become sharper and more pronounced, so that they can be identified up to $n=9$, as indicated in  Fig.~\ref{fig:PerfectSpec}(b). 
Panel (c) of Fig.~\ref{fig:PerfectSpec} focuses on the high-$n$ region, i.e. from $n=10$ to the band gap. Again, the increase of peak absorption at $P$ exciton resonance energies becomes obvious. For the states from $n=11$ to $n=15$ the effect is most prominent. For these states the peak absorption increases by a factor of 2 relative to the background. This is indicated by the vertical blue and red dashed lines close to the state $n=12$. 
Moreover, in this range of principal quantum numbers, intermediate absorption features in between adjacent $P$ exciton lines are observed in the unpumped spectrum, that may consist of  coherent superpositions of adjacent exciton lines~\cite{grunwaldSignaturesQuantumCoherences2016} or even parity $D$ excitons~\cite{lynchRydbergExcitonsSynthetic2021}. A final clarification about the origin of these features is beyond the scope of this manuscript and will be provided in a follow-up investigation. 
Along with the general reduction of absorption in between the $P$ exciton lines, these structures disappear as well for all $n$ below $n=16$. 
Remarkably, the total number of visible $P$ exciton lines does not change.

\begin{figure}
	\begin{center}
		\includegraphics[width=\columnwidth]{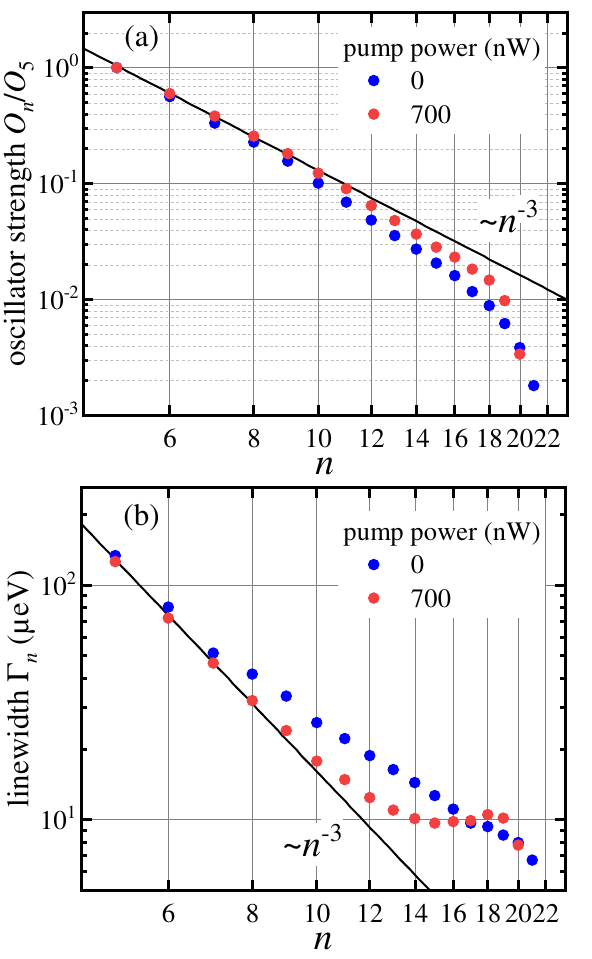} 
	\end{center}
	\caption{Normalized oscillator strengths (a) and linewidths (b) from the spectra in Fig.~\ref{fig:PerfectSpec} at zero pump power (blue) and at 700 nW (red) obtained by fits according to Eq.~\eqref{Toyozawa}. 
		In both panels the black solid lines shows the expected dependence following $n^{-3}$. }
	\label{fig:Scaling}
\end{figure}

The change of absorption induced by the pump laser can be analyzed more quantitatively by fitting asymmetric Lorentzians to the spectrum~\cite{toyozawaInterbandEffectLattice1964, uenoContourAbsorptionLines1969}. 
The asymmetric Fano-like lineshape is described as 
\begin{equation}\label{Toyozawa}
\alpha_n(E)=\frac{O_n}{\pi}\frac{\Gamma_n/2+2q_n(E-E_n)}{\left( \Gamma_n/2\right)^2+(E-E_n)^2} \ , 
\end{equation}
with the oscillator strength $O_n$, the linewidth $\Gamma_n$, the asymmetry parameter $q_n$ and the resonance energy $E_n$.
In Fig.~\ref{fig:Scaling}, we compare the resulting oscillator strengths (panel a) and linewidths (panel b) of the unpumped spectrum (blue in Fig.~\ref{fig:PerfectSpec}) to the values obtained from the pumped spectrum (red curve in Fig.~\ref{fig:PerfectSpec}). 
The trends expected from the scaling laws for an ideal system are shown by the black solid lines. For the oscillator strengths one expects a scaling with the principal quantum number as $n^{-3}$ according to Elliott's theory~\cite{elliottIntensityOpticalAbsorption1957}. The linewidths are determined by both radiative decay and phonon scattering, which also both result in a scaling as $n^{-3}$~\cite{gallagherRydbergAtoms1994, stolzInteractionRydbergExcitons2018}. 
In the unpumped spectrum (blue circles), the values for both parameters start to deviate from the expected behavior at around $n=8$. The oscillator strengths are smaller than expected and drop even stronger towards higher principal quantum numbers. The linewidths decrease with $n$, but less strong than expected from a pure $n^{-3}$ scaling. 
The line parameters obtained from the pumped spectrum (red dots)  follow the predicted scalings much more closely. Up to $n=10$ they match almost perfectly with the ideal values. 
However, for states with principal quantum numbers larger than $n=10$ deviations from the ideal values can still be observed. In particular, the linewidths lose their dependence on $n$ and start to saturate around a width of about 10~$\mu$eV in the high-$n$ regime, see Fig.~\ref{fig:Scaling}(b). 

The deviations of the line shape parameters from ideal scalings were shown to stem from the presence of charged impurities in the crystal in Ref.~\cite{krugerInteractionChargedImpurities2020}. Accordingly, the pump laser induced reduction of these deviations is interpreted as a reduction of the amount of charged impurities in the crystal. We call this effect \textit{purification}, and will discuss it further in the following sections.

\subsubsection{Purification: Pump power dependence}\label{power}
\begin{figure}
	\begin{center}
		\includegraphics[width=\columnwidth]{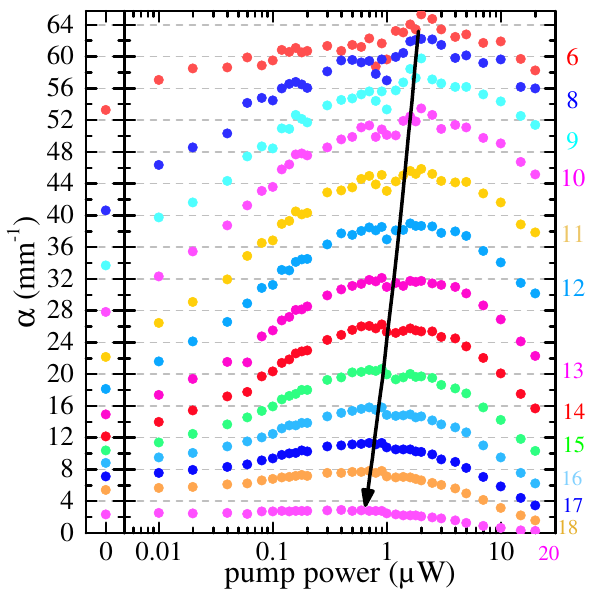} 
	\end{center}
	\caption{Exciton peak absorption as a function of pump power for states from $n_\text{probe}=6$ to $n_\text{probe}=20$. The probe power is set to 200~nW, the pump laser energy is set close to $\tilde{E}_\text{g}$. For all $n_\text{probe}$ the absorption increases, while the power at which the maximal increase is obtained varies from 2~$\mu$W at low $n$ (top) to 700~nW at high $n$ (bottom), as indicated by the arrow. The absorption increases by more than a factor of 2 for the states $n=11$ to 15, compared to the value at zero pump power shown in the left panel.  The data is obtained by careful subtraction of the continuous background (see vertical dashed lines in Fig.~\ref{fig:PerfectSpec}(c)). 
	 }
	\label{fig:Scaling_height}
\end{figure}

The data shown  so far was measured at pump laser power of 700~nW. This gives rise to the question if the effect of purification becomes even stronger at higher powers. Therefore, we repeat the measurement shown in Fig.~\ref{fig:PerfectSpec} for various pump powers and analyze the change of exciton peak absorption as a function of pump power. The exciton peak absorption is extracted from the spectra by careful subtraction of the continuous background, as indicated by the vertical blue and red lines in Fig.~\ref{fig:PerfectSpec}(c) next to $n=12$. Fig.~\ref{fig:Scaling_height} shows the change of peak absorption for all excitons from $n_\text{probe}=6$ to 18 as a function of pump laser power. 
For all probed $n$ up to $n_\text{probe}=18$, the absorption increases strongly with increasing pump power until it reaches a maximum. At higher powers, the absorption drops again indicating the onset of blockade effects as mentioned before. 
The pump power at which a maximal probe absorption is reached varies slightly with  $n_\text{probe}$. It is found to be around 2~$\mu$W for low $n$ and around 700~nW for the highest $n$ around $n=18$, as indicated by the arrow in Fig.~\ref{fig:Scaling_height}. 
In general, the powers at which  maximal purification is reached  in sample I are low compared to pump powers used in earlier reports focusing on Rydberg blockade or exciton-plasma interaction~\cite{heckotterAsymmetricRydbergBlockade2021, heckotterRydbergExcitonsPresence2018}. 
  The absolute increase in absorption compared to the unpumped case is strongest for excitons between $n=11$ to 15 and amounts to more than a factor of 2 compared to the value at zero pump power. Remarkably, pump powers of 100 nW are already sufficient to obtain growth by 50~\%. 

\begin{figure}
	\begin{center}
		\includegraphics[width=\columnwidth]{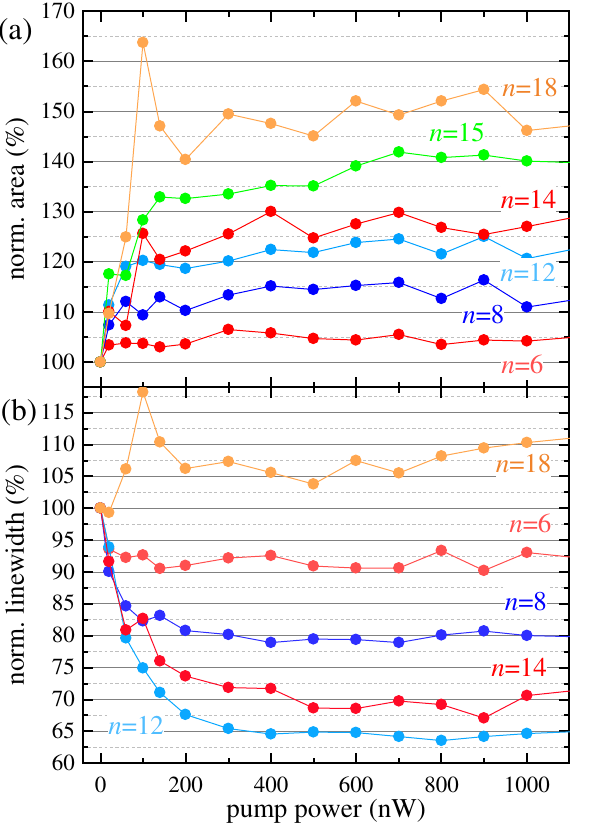} 
	\end{center}
	\caption{Analysis of (a) oscillator strengths $O_n$ and (b) linewidths $\Gamma_n$ as a function of pump power for even principal quantum numbers measured at a probe power of 200~nW. The pump laser energy is set to $\tilde{E}_\text{g}$. The pump power range is restricted to the pump power range of purification, i.e. below 2~$\mu$W. The values are normalized to the corresponding value at zero pump power to visualize the change in percent. }
	\label{fig:Scaling_linewidth_osc}
\end{figure}

This enhancement of exciton peak absorption, measured at the resonance energy $E_n$, may be directly connected to an increase of oscillator strength or a narrowing of linewidth $\alpha_n(E_n)\propto \frac{O_n}{\Gamma_n}$. 
To access these two parameters independently, we analyze  the data by fits according to Eq.~\eqref{Toyozawa}. 
Both $O_n$ and $\Gamma_n$ are shown in Fig.~\ref{fig:Scaling_linewidth_osc} for even principal quantum numbers as a function of pump laser power, relative to their unpumped value. Indeed, the oscillator strengths increase and the linewidths narrow up to a power of 700~nW. The strongest increase of oscillator strength  by up to 50~\% is found for $n=18$. For the linewidths, the strongest drop is found for $n=12$, which decreases by about 35~$\%$ already at pump powers of 400~nW. 

\subsubsection{Purification: Pump laser energy dependence}\label{wavelength}
\begin{figure*}[!htb]\label{pumpscanall}
	\begin{center}
		\includegraphics[width=\textwidth]{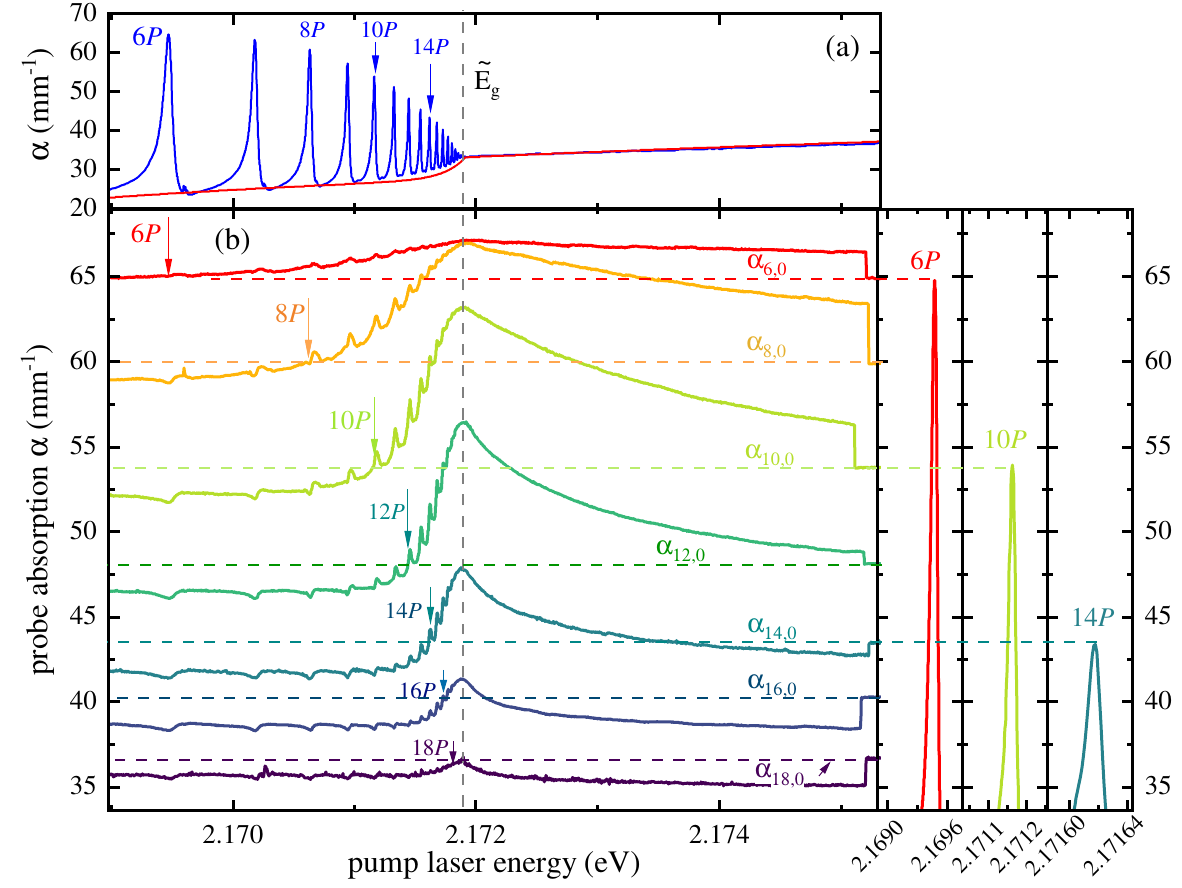} 
	\end{center} 
	\caption{(a) Linear absorption spectrum. Data taken from \cite{stolzCoherentTransferMatrix2021}. The red line describes the background that consists of a phonon absorption and an exponential-like tail below $\tilde{E}_g$.
		(b) Probe laser absorption at excitonic resonances with different $n_\text{probe}$ as a function of pump laser  energy. For all $n_\text{probe}$ the enhancement of absorption is maximal close to $\tilde{E}_\text{g}$. The presence of the pump laser reduces or increases the probe laser absorption depending drastically on its energy. The window of pump laser energies that cause an enhanced absorption decreases with increasing $n_\text{probe}$.  The absorption in absence of the pump laser is given by the plateau in the right corner, where the pump laser is blocked. This plateau corresponds to the absorption on resonance in the unpumped case, $\alpha_{n,0}$, as shown in the right panels (see dashed horizontal lines).  }
	\label{fig:pumpscannew}
\end{figure*}

To investigate the effect of purification in more detail, we discuss its dependence on pump laser energy in the following. 
Therefore, we fix the probe laser energy on specific resonances and scan the pump laser's energy $E_\text{pump}$. A similar measurement is discussed in Ref.~\cite{bergenTamingChargedDefects2023} for a single probe energy. In the present manuscript, we extend the investigation of Ref.~\cite{bergenTamingChargedDefects2023} by setting the probe laser to every second resonance from $n_\text{probe}=6$ to $18$. The probe (pump) laser is set to a power of 1 (10) $\mu$W. 
As a reference to the relevant  range of pump laser energies, we show a linear absorption spectrum of the crystal in panel (a) of Fig.~\ref{fig:pumpscannew}. 
The red dashed line describes the underlying background absorption as a guide to the eye. 
It consists of the phonon background as well as the continuum absorption and, in particular, an exponential Urbach-like tail $\sim \exp((E-\tilde{E}_\text{g})/E_U)$ below the band gap~\cite{stolzScrutinizingDebyePlasma2022,schonePhononassistedAbsorptionExcitons2017}. Here, $E_U=170~\mu$eV is used for the width of the tail. 
This exponential tail results from the overlap of broadened Rydberg exciton lines close to the band gap~\cite{krugerInteractionChargedImpurities2020}. 

Fig.~\ref{fig:pumpscannew}(b) shows the  probe  laser absorption as a function of pump laser energy for all probed states. 
The probe absorption in the absence of the pump laser, $\alpha_{n,0}$, is indicated by the horizontal dashed lines. It corresponds to the absorption of the unpumped spectrum, shown exemplarily in the right panels for $n=6, 10$ and $14$ and is obtained by blocking the pump laser (see plateaus at the high-energy side of each curve). 

For all $n_\text{probe}$, the probe laser absorption shows the same characteristic dependence on the pump laser energy:
At low pump energies, the pump laser causes a reduction of probe absorption compared to $\alpha_{n,0}$. 
The trend of the probe absorption  mirrors the pump absorption, as it consists of a continuous trend that is superimposed by discrete dips at energies that a match with the exciton resonance energies in panel (a).  
At these energies the pump laser light may be absorbed into low $nP$ exciton states and indirectly into $1S$ excitons via the phonon background. 
Hence, Rydberg blockade~\cite{heckotterAsymmetricRydbergBlockade2021} or a low-density of an electron-hole plasma, created via Auger-decay of $1S$ excitons~\cite{stolzScrutinizingDebyePlasma2022} may reduce the probe absorption. 
As discussed in Ref.~\cite{bergenTamingChargedDefects2023}, the exact composition of the excitations created by the pump laser is complex and may also vary with pump laser energy. 
A final clarification is beyond the scope of this manuscript. In this regard, time-resolved measurements may resolve the different dynamics of the underlying processes and give better access compared to cw-spectra.

As the pump energy increases, the reduction of probe absorption at low pump energies becomes superimposed by an counteracting continuous increase that starts roughly around the energy of $n=6$. 
However, an enhancement of absorption - or purification - , i.e.  $\alpha_n>\alpha_{n,0}$, is solely observed for pump energies equal or higher than the probe laser energy, $E_\text{pump}\geq E_\text{probe}$. 
This becomes obvious by comparing the horizontal lines with the vertical arrows.  
Hence, the onset of the spectral range of pump laser energies that purify an exciton line shifts to higher energies with rising $n_\text{probe}$. 

The increasing probe laser absorption forms an exponential tail at pump energies below the band gap $\tilde{E}_g$. The tail is superimposed by discrete peaks. Also here, the probe laser absorption mirrors the absorption spectrum in panel (a) that shows prominent $nP$ absorption lines and the exponential absorption tail (red line). 
 We conclude that purification is directly connected to the excitation of Rydberg excitons excited either resonantly or via the exponential-like tail close to the band gap $\tilde{E}_\text{g}$. 
 For pump laser energies in resonance with a Rydberg exciton line the enhancement is pronounced compared to the exponential tail, as the density of excitons created is larger.

For all $n_\text{probe}$, the pump laser energy corresponding to the maximal enhancement is found close to the apparent band gap $\tilde{E}_\text{g}$. 
It varies only within 40~$\mu$eV  among all probed $n$ from $n_\text{probe}=8$ to 18. An exception is found for $n_\text{probe}=6$, where the energy of maximal enhancement is shifted by about 90~$\mu$eV above $\tilde{E}_\text{g}$. 

When extending the pump energy beyond the band gap, the enhanced absorption diminishes again. Nevertheless, purification can be observed for pump energies above the band gap as well, whereas the spectral range strongly decreases with increasing $n_\text{probe}$. 
At these energies, the excitation may initially create a low-energy electron-hole plasma, which  causes purification directly or after relaxation into high-$n$ Rydberg states. 
The free charges may also reduce the absorption by plasma-screening effects~\cite{heckotterRydbergExcitonsPresence2018} and counteract the enhanced absorption. This counteracting process then becomes stronger for higher $n$. 
The total spectral pump range that results in enhanced probe laser absorption depends drastically on $n_\text{probe}$, decreasing from lower to higher states. For $n_\text{probe}=16$ it is as narrow as 350~$\mu$eV and reduces to around 90~$\mu$eV for $n_\text{probe}=18$.

\begin{figure}[h]
	\begin{center}
		\includegraphics[width=1\columnwidth]{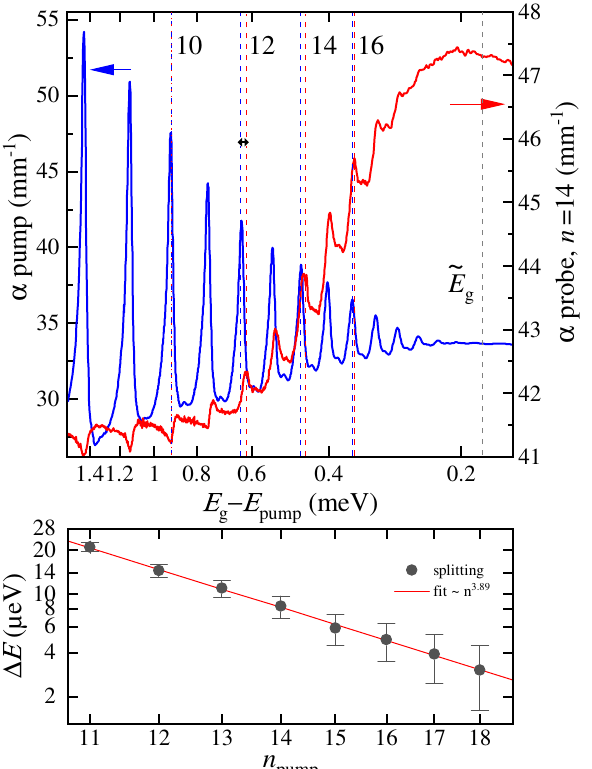} 
	\end{center}
	\caption{  
		(a) Direct comparison of the pump (left axis) and probe (right axis) laser absorption in the spectral below the band gap. The energy is given relative to the band gap $E_\text{g}$ on a logarithmic axis. 
		At  $n_\text{pump}=10$ , both maximal pump laser absorption and maximally reduced probe laser absorption occur at the same pump laser energy (vertical dashed line). 
		From $n_\text{pump}=11$ onwards, the maximal enhancement of probe laser absorption (red dashed line) is found at laser energies slightly above the energy of maximal pump absorption (blue dashed line). 
		(b) The splitting between the energies of maximal probe enhancement and maximal pump absorption as a function of pump $n_\text{pump}$. The red line shows a fit according to $\sim n_\text{pump}^{-3.89}$. 
	}
	\label{fig:PumpScan2}
\end{figure}

In order to separate the effect of resonant exciton excitation  from absorption into the continuous background, we compare the data from Fig.~\ref{fig:pumpscannew}(a) and (b) in Fig.~\ref{fig:PumpScan2}(a) directly in a smaller energy range. Note the logarithmic energy axis. 
Pumping at low $n_\text{pump}$, e.g. $n_\text{pump}=10$, induces a reduction of probe laser absorption relative to pumping in the continuous background. 
A comparison with the linear  absorption spectrum (blue) reveals that the magnitude of the pump absorption directly translates into the response of the probe laser: Hence, both the maximal reduction of probe laser absorption and the maximal pump absorption occur when the pump energy matches the exciton peaks, as indicated by the vertical dashed lines at $n_\text{pump}=10$. 
For higher $n_\text{pump}$ the pump laser instead induces an enhancement of probe laser absorption relative to the continuous background (purification). Here, the situation is different: The maximal enhancement is observed at pump laser energies slightly above the energy of maximal pump absorption, as indicated by the separated blue and red vertical lines. 
The splitting between these energies is plotted in Fig.~\ref{fig:PumpScan2}(b) versus the principal quantum number $n_\text{pump}$. The splitting drops with increasing $n_\text{pump}$. A fit to the data reveals a dependence on the principle quantum number as $ n_\text{pump}^{-3.89}$, which is close to a scaling law of $n^{-4}$. 

\subsection{Sample of inferior quality} \label{sample2}
\begin{figure}
	\begin{center}
		\includegraphics[width=\columnwidth]{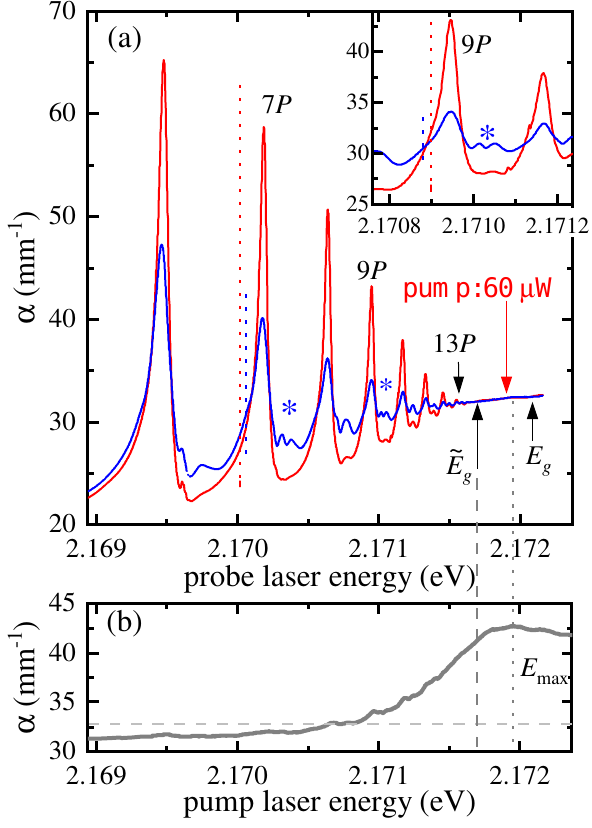} 
	\end{center}
	\caption{Purification at sample II of inferior quality. 
		(a) The bare absorption spectrum  is drawn in blue showing strongly decreasing $P$ exciton peaks up to $n_\text{max}=14$ and pronounced absorption features in between the $P$ exciton peaks (blue stars).
		The spectrum with additional pump laser is drawn in red. The pump laser induces an enhancement of absorption of around a factor of 2 for $n=7$ and a factor of 3 for $n=10$ (inset). The absorption in between the $P$ excitons is reduced. 
		(b) Probe laser absorption at $n=9$ as a function of pump laser energy. The pump laser energy for maximal purification is found to be at $E_\text{max}=2.17195$~eV. The grey dashed horizontal line indicates the absorption  of $n_\text{probe}=9$ without perturbation by the pump laser. For this scan the pump laser power was set to 40~$\mu$W.  }
	\label{fig:badsample}
\end{figure}

The reported enhancement of absorption of exciton peaks was shown for sample I, a natural sample of high quality. In this Section, we study the effect of purification on sample II, an artificially grown gemstone of inferior quality, i.e. a sample with a larger amount of impurities. 
A high resolution absorption spectrum is shown by the blue curve in Fig.~\ref{fig:badsample}(a) starting from $n=6$. In this sample, the Rydberg series ends at $n_\text{max}=14$ and the absorption strength of the visible $P$ states is reduced compared to sample I (cf. Fig.~\ref{fig:PerfectSpec}). 
Moreover, additional absorption features can be observed in between the $P$ lines, which are more prominent compared to sample I and even appear split into two lines at some particular $n$, as indicated by the blue stars. 

We identify the pump laser energy where the maximal amplitude of purification occurs by performing a scan of the pump laser energy, similar to Fig.~\ref{fig:pumpscannew}. It  is shown in panel (b) of Fig.~\ref{fig:badsample}. 
Remarkably, the pump laser energy at which the increase of absorption takes its maximum is found to be $E_\text{max}=2.17195$~eV, as indicated by the grey dotted line. 
Compared to the spectrum in panel (a), this energy is about 250~$\mu$eV higher than the apparent band gap $\tilde{E}_\text{g}=2.17170$~eV, which is defined here as the high-energy side of $n_\text{max}=14$. 

Following the interpretation given above, that the enhancement of absorption is caused by the excitation of Rydberg excitons, either resonantly or non-resonantly via absorption into the exponential absorption tail, we conclude that we still excite Rydberg excitons at energies 250~$\mu$eV above the apparent gap $\tilde{E}_\text{g}$, although no sharp absorption lines are visible. 
Hence, the observed flat continuum may be formed by a complex overlap of broadened Rydberg exciton lines. 

With the additional pump laser exciting close to the energy $E_\text{max}$, we achieve the maximal purification in this sample at a power of 60 $\mu$W. The resulting spectrum is shown in red in Fig.~\ref{fig:badsample}(a). 
The change of the absorption spectrum is even more prominent compared to sample I: 
The absorption  doubles at $n=7$ and increases almost a factor of 3 for $n=10$ in this sample. 
For clarity, the exciton absorption strength relative to the background is indicated  by the vertical red and blue dashed lines for $n=7$ and $n=10$ (inset) for the cases with and without pump laser. 

The optimal power for purification of 60~$\mu$W is roughly 60 times higher compared to the value found for the high-quality sample I. This underlines the assumption of a larger total amount of charged impurities in this sample, as a higher density of Rydberg excitons is necessary to reach the maximally purified state. 
The absorption features in between the $P$ exciton lines decrease for all $n$ (blue stars), while they disappear completely only for lower $n$. 
Assuming that the features in between the $P$ excitons stem from even parity excitons that gain oscillator strength due to electric field-induced mixing with $P$ excitons caused by charged impurities, the visibility of those features even at the saturation power of 60~$\mu$W indicates that not all defects can be fully neutralized by applying the additional pump laser beam. 
Finally, the highest observable principal quantum number does not change, similar to sample I.

\section{Impurity luminescence}\label{PL}
The absorption spectra shown in the previous sections imply that Rydberg excitons interact with charged defects and their presence may change the defect's effective charge. 
To underspin this assumption, we investigate the impurity-related PL signals in sample I under additional laser excitation. 

In Cu$_2$O, vacancy emission is typically observed at energies below the 1$S$ exciton transition, i.e. below 2.03~eV~\cite{koiralaRelaxationLocalizedExcitons2014}. 
In natural crystals, typically two major signals from excitons bound to singly or doubly charged O$^+$ and O$^{++}$ vacancies are observed, while Cu$^+$ defects are more prominent in artificially grown crystals~\cite{frazerVacancyRelaxationCuprous2017, lynchRydbergExcitonsSynthetic2021}. 
At energies above the O$^{++}$ vacancy, typically phonon-assisted emission lines of 1$S$ para- and ortho excitons are observed~\cite{takahataPhotoluminescencePropertiesEntire2018a}. Moreover, in natural samples also a broad emission feature is observed in this spectral region, commonly interpreted in terms of excitons bound to metallic impurities~\cite{jangBoundExcitonsMathrmCu2006, itoDetailedExaminationRelaxation1997,steinhauerRydbergExcitonsCu2O2020}. 

\begin{figure}
	\begin{center}
		\includegraphics[width=\columnwidth]{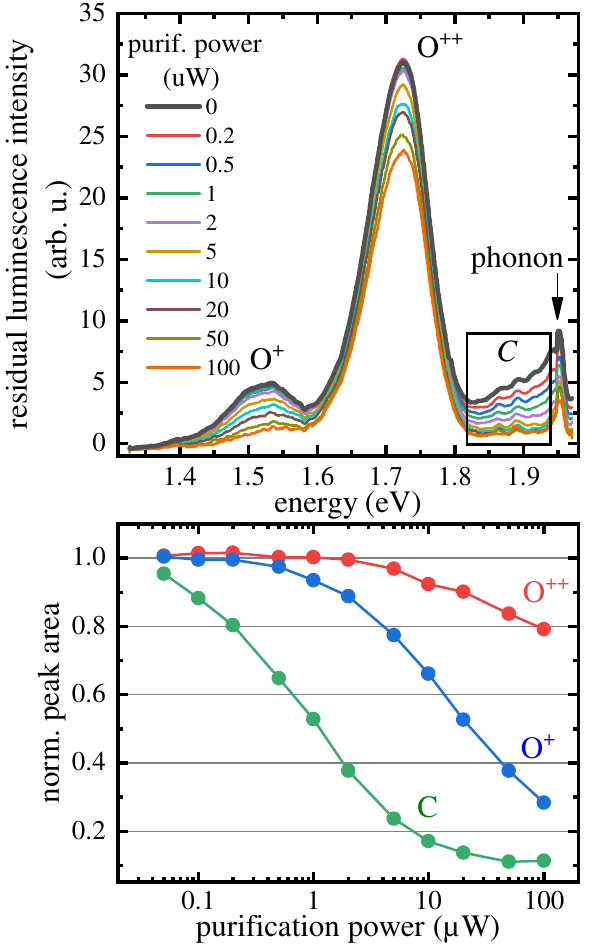} 
	\end{center}
	\caption{Residual PL intensity of defects in sample I under above-band-gap excitation at 2.033 eV and additional \textit{purification} laser excitation at $\tilde{E}_\text{g}$ with varying powers. Both $O^+$, $O^{++}$ as well as the signal in range \textbf{C} decrease with increasing power of the additional laser. 
	(b) Peak areas from the signals in panel (a) obtained by a fit (see text) normalized to the value in the absence of additional laser excitation. } 
	\label{fig:PLPower}
\end{figure}

First, we excite the crystal non-resonantly at 2.33~eV (532 nm) with a power of 10~$\mu$W. 
After creation of free carriers by the non-resonant laser excitons may form, which in turn become trapped at the impurities. Radiative recombination then results in prominent emission features for each type of impurity~\cite{koiralaRelaxationLocalizedExcitons2014}. The resulting PL spectrum is shown in Fig.~\ref{fig:PLPower} by the bold black line. 
In agreement with the descriptions above, the PL spectrum contains both singly and doubly charged oxygen peaks, $O^+$ and $O^{++}$, at 1.52 and 1.72~eV, as well as a broad emission feature around 1.9~eV, that is indicated with a \textbf{C}. Due to its broad character, we attribute the latter signal to excitons bound to metallic impurities. 
Phonon replica typically give rise to several discrete peaks at energies above 1.95~eV~\cite{takahataPhotoluminescencePropertiesEntire2018a}. In Fig.~\ref{fig:PLPower}, the lowest phonon replica, stemming from the $\Gamma_4^-$ (TO) phonon, can be observed as a sharp signal at 1.95~eV, as indicated. 

Next, we add a second laser with a photon energy fixed at $\tilde{E}_\text{g}$, which we refer to as \textit{purification laser} in order to distinguish it from the first one. 
We analyze the change in the PL spectrum under additional excitation by this laser. 
As we detect the emission created by the \textit{purification laser} itself  as well, it is carefully subtracted from the total signal to obtain the residual PL signal created solely by the non-resonant laser.   
Using this procedure, any sub-linear dependence of the total PL emission~\cite{jangBoundExcitonsMathrmCu2006}  on total excitation power will also result in changes of the residual PL signal. 
	We carefully checked that the induced changes by the \textit{purification laser} are stronger than expected from the sub-linear power dependence of the emission features. 
	
To determine the intensity of each kind of impurity signal, the oxygen peaks are fitted by Gaussians with amplitudes $a$ 
and widths $\sigma$ 
at energies $E$ 
, while the emission $C$ is approximated by an exponential background decaying from $E_0=1.95$~eV to lower energies, 
with a total amplitude $a_C$ and a parameter $\sigma_C$ accounting for the decay strength.  The total fitting function then reads
$a_C\cdot e^{-\sigma_C(E-E_0)}+ a_+\cdot e^{-\left(\frac{E-E_+}{\sigma_+}\right)^2} + a_{++}\cdot e^{-\left(\frac{E-E_{++}}{\sigma_{++}}\right)^2}$ \. 
As shown in Fig.~\ref{fig:PLPower}(b), the obtained intensities of these three impurity signals in the residual PL drop with increasing power of the purification laser. 
The signal in range \textbf{C} is reduced by about 80~\% at 10~$\mu$W.  

The total intensity of the non-resonantly excited PL signal is proportional to the amount of singly or doubly charged oxygen vacancies or  metallic defects present in the material and an additional laser set to an energy close to the band gap reduces this part of the PL. 
Hence, this drop of PL intensity is interpreted as a reduction of the amount of charged impurities caused by the purification laser in agreement with the observations reported above. Again, the reduction stems from pump-induced excitons that neutralize or screen a part of charged impurities in the material.

\section{Conclusion}\label{discussion}
In this work, we reported on a laser-induced enhancement of absorption in the $P$ exciton spectrum of Cu$_2$O. We showed that a low excitation power of only 700~nW is sufficient to improve the overall quality of the absorption spectrum in terms of an increase of oscillator strengths and narrowing of linewidths. For excitons with principal quantum number $n=11$ to 15, this results in a doubling of peak absorption. 
Furthermore, $F$-excitons with a high angular momentum $L=3$ become visible at principal quantum numbers up to $n=9$. 

This {purification} of the crystal is observed for additional laser excitation within a narrow spectral window around the apparent band gap $\tilde{E}_\text{g}$. 
It starts to appear as an exponential increase at energies below the band gap, where the pump laser excites Rydberg excitons with highest principal quantum numbers available, either resonantly or via absorption into the exponential absorption tail, that is formed by the complex overlap of broadened Rydberg-exciton lines. An effective enhancement of the probe laser absorption is observed solely for pump laser energies that are larger than the probed state's  energy, i.e. $E_\text{pump}\geq E_\text{probe}$. 
This is in agreement with the observations under pulsed excitation reported in Ref.~\cite{bergenTamingChargedDefects2023}. While the model presented there focuses on the effect of charged impurities, it could be extended by taking other excitations into account that become relevant under cw excitation, like an accumulation of $1S$ excitons and a formation of an electron-hole plasma. 
	
The maximal effect of {purification} is observed at pump laser energies close to $\tilde{E}_\text{g}$. 
Since charged impurities in Cu$_2$O are known to reduce the quality of exciton absorption spectra, the reported observations are interpreted as a neutralization or screening of charged impurities by the injection of Rydberg excitons. 

Even at pump laser energies above the band gap an enhancement is observed, whereby the spectral range strongly depends on the probed state. Here, a low-energy plasma is excited that may screen impurities directly or relax into a complex distribution of high-$n$ Rydberg excitons. 
The spectral window of pump laser energies that purifies an exciton line becomes narrower with increasing $n$ of the probed exciton state. 

In the high-quality sample I, the purification of the crystal saturates already at low pump powers below 1~$\mu$W. At these low powers, the small density of Rydberg excitons is sufficient to neutralize a large fraction of defects. At higher powers, blockade effects set in, that overlay the purification again. 
In this line, we found a much higher  power at which the maximal purification of the crystal is reached in the low-quality sample II. 

According to Ref.~\cite{krugerInteractionChargedImpurities2020}, the density of impurities defines the maximum principle quantum number $n_\text{max}$. However, in both samples we do not observe an increase of the maximum principal quantum number $n_\text{max}$ in presence of the pump laser. For high-$n$ states, purification becomes almost negligible. We think that due to the presence of the pump laser also free charges may be    created, either by the neutralization process itself or by additional absorption channels that counteract the purification at high-$n$ states. 

Our interpretation of a change of the charged impurity density in the material by additional illumination is supported by investigations of the impurity PL signals. We observed a reduction of residual impurity emission when an additional purification laser close to the energy $\tilde{E}_\text{g}$ is added. 

The microscopic mechanism behind the capture process of Rydberg excitons at charged defects is beyond the scope of the present manuscript. In a cw-excitation scheme as used here, the indirect formation of an electron-hole plasma  by Auger-decay of 1$S$ excitons cannot be  excluded. Moreover, the direct excitation of Rydberg-excitons also introduces changes in the 	absorption mediated by Rydberg blockade. Both effects might overlay and counteract the observed purification which complicates an exact quantitative analysis. 
	However, these effects can be separated using pulsed excitation and a time-resolved detection scheme, see Ref.~\cite{bergenTamingChargedDefects2023}. 

\section{Outlook}\label{outlook}
Further investigations may aim on neutralization of impurities under different experimental conditions, e.g. under the presence of an electron-hole plasma or at higher temperatures, where impurities are screened or thermally ionized. 

We propose to utilize this technique of {purification} in general for future experiments in order to cancel out the detrimental impact of defects on excitons  and to create an as-pure-as-possible environment before starting experiments on all kinds of physical questions. 

The observed narrowing of lines could also improve even-exciton spectra in Cu$_2$O, as obtained by two-photon absorption or second-harmonic generation (SHG) techniques, where the experimentally observed lines are typically rather broad~\cite{mundHighresolutionSecondHarmonic2018a, rogersHighResolutionNanosecond2021}. 
Also, recently reported optical non-linearities~\cite{morin_self-kerr_2022} as well as intra-exciton transitions  \cite{gallagher_microwave-optical_2022} may appear stronger in a purified Cu$_2$O crystal. 

The technique of purification can be used to control the amount of ionized impurities and to improve the quality of absorption spectra even for artificially grown samples of lower quality, as we showed in Sec.~\ref{sample2}.
While we focused on bulk samples with a thickness of tens of $\mu$m here, this technique might also improve the quality of  micro- and nanocrystals or thin films~\cite{steinhauerRydbergExcitonsCu2O2020, orfanakisQuantumConfinedRydberg2021, nakaThinFilmsSingleCrystal2005}. 
Moreover, in a sample of lower quality we found purification to be most efficient in a spectral range, where no distinct exciton peaks are observed in the  absorption spectrum (Fig.~\ref{fig:badsample}). Hence, a similar improvement might be obtained  also in other systems like 2D materials, where excitonic Rydberg series suffer from broadened lines and where no well defined resonances are observable in the proximity of the band gap~\cite{chernikov_exciton_2014,hill_observation_2015}. 
Therefore, our finding is of general importance for  semiconductor-based quantum technologies.

\begin{acknowledgments}
	We would like to thank D. Fr\"ohlich for fruitful discussions and for providing the artificial sample. 
We acknowledge the financial support by the Deutsche Forschungsgemeinschaft through the project 504522424. 
\end{acknowledgments}


\begin{thebibliography}{57}%
\makeatletter
\providecommand \@ifxundefined [1]{%
 \@ifx{#1\undefined}
}%
\providecommand \@ifnum [1]{%
 \ifnum #1\expandafter \@firstoftwo
 \else \expandafter \@secondoftwo
 \fi
}%
\providecommand \@ifx [1]{%
 \ifx #1\expandafter \@firstoftwo
 \else \expandafter \@secondoftwo
 \fi
}%
\providecommand \natexlab [1]{#1}%
\providecommand \enquote  [1]{``#1''}%
\providecommand \bibnamefont  [1]{#1}%
\providecommand \bibfnamefont [1]{#1}%
\providecommand \citenamefont [1]{#1}%
\providecommand \href@noop [0]{\@secondoftwo}%
\providecommand \href [0]{\begingroup \@sanitize@url \@href}%
\providecommand \@href[1]{\@@startlink{#1}\@@href}%
\providecommand \@@href[1]{\endgroup#1\@@endlink}%
\providecommand \@sanitize@url [0]{\catcode `\\12\catcode `\$12\catcode
  `\&12\catcode `\#12\catcode `\^12\catcode `\_12\catcode `\%12\relax}%
\providecommand \@@startlink[1]{}%
\providecommand \@@endlink[0]{}%
\providecommand \url  [0]{\begingroup\@sanitize@url \@url }%
\providecommand \@url [1]{\endgroup\@href {#1}{\urlprefix }}%
\providecommand \urlprefix  [0]{URL }%
\providecommand \Eprint [0]{\href }%
\providecommand \doibase [0]{https://doi.org/}%
\providecommand \selectlanguage [0]{\@gobble}%
\providecommand \bibinfo  [0]{\@secondoftwo}%
\providecommand \bibfield  [0]{\@secondoftwo}%
\providecommand \translation [1]{[#1]}%
\providecommand \BibitemOpen [0]{}%
\providecommand \bibitemStop [0]{}%
\providecommand \bibitemNoStop [0]{.\EOS\space}%
\providecommand \EOS [0]{\spacefactor3000\relax}%
\providecommand \BibitemShut  [1]{\csname bibitem#1\endcsname}%
\let\auto@bib@innerbib\@empty
\bibitem [{\citenamefont {Kazimierczuk}\ \emph {et~al.}(2014)\citenamefont
  {Kazimierczuk}, \citenamefont {Fröhlich}, \citenamefont {Scheel},
  \citenamefont {Stolz},\ and\ \citenamefont
  {Bayer}}]{kazimierczukGiantRydbergExcitons2014c}%
  \BibitemOpen
  \bibfield  {author} {\bibinfo {author} {\bibfnamefont {T.}~\bibnamefont
  {Kazimierczuk}}, \bibinfo {author} {\bibfnamefont {D.}~\bibnamefont
  {Fröhlich}}, \bibinfo {author} {\bibfnamefont {S.}~\bibnamefont {Scheel}},
  \bibinfo {author} {\bibfnamefont {H.}~\bibnamefont {Stolz}},\ and\ \bibinfo
  {author} {\bibfnamefont {M.}~\bibnamefont {Bayer}},\ }\bibfield  {title}
  {\bibinfo {title} {Giant {Rydberg} excitons in the copper oxide
  $\mathrm{Cu}_2\mathrm{O}$},\ }\href {https://doi.org/10.1038/nature13832}
  {\bibfield  {journal} {\bibinfo  {journal} {Nature}\ }\textbf {\bibinfo
  {volume} {514}},\ \bibinfo {pages} {343} (\bibinfo {year}
  {2014})}\BibitemShut {NoStop}%
\bibitem [{\citenamefont {Schweiner}\ \emph
  {et~al.}(2017{\natexlab{a}})\citenamefont {Schweiner}, \citenamefont {Main},
  \citenamefont {Wunner}, \citenamefont {Freitag}, \citenamefont
  {Heck{\"o}tter}, \citenamefont {Uihlein}, \citenamefont {A{\ss}mann},
  \citenamefont {Fr{\"o}hlich},\ and\ \citenamefont
  {Bayer}}]{schweinerMagnetoexcitonsCuprousOxide2017a}%
  \BibitemOpen
  \bibfield  {author} {\bibinfo {author} {\bibfnamefont {F.}~\bibnamefont
  {Schweiner}}, \bibinfo {author} {\bibfnamefont {J.}~\bibnamefont {Main}},
  \bibinfo {author} {\bibfnamefont {G.}~\bibnamefont {Wunner}}, \bibinfo
  {author} {\bibfnamefont {M.}~\bibnamefont {Freitag}}, \bibinfo {author}
  {\bibfnamefont {J.}~\bibnamefont {Heck{\"o}tter}}, \bibinfo {author}
  {\bibfnamefont {C.}~\bibnamefont {Uihlein}}, \bibinfo {author} {\bibfnamefont
  {M.}~\bibnamefont {A{\ss}mann}}, \bibinfo {author} {\bibfnamefont
  {D.}~\bibnamefont {Fr{\"o}hlich}},\ and\ \bibinfo {author} {\bibfnamefont
  {M.}~\bibnamefont {Bayer}},\ }\bibfield  {title} {\bibinfo {title}
  {Magnetoexcitons in cuprous oxide},\ }\href
  {https://doi.org/10.1103/PhysRevB.95.035202} {\bibfield  {journal} {\bibinfo
  {journal} {Physical Review B}\ }\textbf {\bibinfo {volume} {95}},\ \bibinfo
  {pages} {035202} (\bibinfo {year} {2017}{\natexlab{a}})}\BibitemShut
  {NoStop}%
\bibitem [{\citenamefont {Rommel}\ \emph {et~al.}(2018)\citenamefont {Rommel},
  \citenamefont {Schweiner}, \citenamefont {Main}, \citenamefont
  {Heck{\"o}tter}, \citenamefont {Freitag}, \citenamefont {Fr{\"o}hlich},
  \citenamefont {Lehninger}, \citenamefont {A{\ss}mann},\ and\ \citenamefont
  {Bayer}}]{rommelMagnetoStarkEffectYellow2018}%
  \BibitemOpen
  \bibfield  {author} {\bibinfo {author} {\bibfnamefont {P.}~\bibnamefont
  {Rommel}}, \bibinfo {author} {\bibfnamefont {F.}~\bibnamefont {Schweiner}},
  \bibinfo {author} {\bibfnamefont {J.}~\bibnamefont {Main}}, \bibinfo {author}
  {\bibfnamefont {J.}~\bibnamefont {Heck{\"o}tter}}, \bibinfo {author}
  {\bibfnamefont {M.}~\bibnamefont {Freitag}}, \bibinfo {author} {\bibfnamefont
  {D.}~\bibnamefont {Fr{\"o}hlich}}, \bibinfo {author} {\bibfnamefont
  {K.}~\bibnamefont {Lehninger}}, \bibinfo {author} {\bibfnamefont
  {M.}~\bibnamefont {A{\ss}mann}},\ and\ \bibinfo {author} {\bibfnamefont
  {M.}~\bibnamefont {Bayer}},\ }\bibfield  {title} {\bibinfo {title}
  {Magneto-{{Stark}} effect of yellow excitons in cuprous oxide},\ }\href
  {https://doi.org/10.1103/PhysRevB.98.085206} {\bibfield  {journal} {\bibinfo
  {journal} {Physical Review B}\ }\textbf {\bibinfo {volume} {98}},\ \bibinfo
  {pages} {085206} (\bibinfo {year} {2018})}\BibitemShut {NoStop}%
\bibitem [{\citenamefont {{Zieli{\'n}ska-Raczy{\'n}ska}}\ \emph
  {et~al.}(2016{\natexlab{a}})\citenamefont {{Zieli{\'n}ska-Raczy{\'n}ska}},
  \citenamefont {Ziemkiewicz},\ and\ \citenamefont
  {Czajkowski}}]{zielinska-raczynskaElectroopticalPropertiesRydberg2016}%
  \BibitemOpen
  \bibfield  {author} {\bibinfo {author} {\bibfnamefont {S.}~\bibnamefont
  {{Zieli{\'n}ska-Raczy{\'n}ska}}}, \bibinfo {author} {\bibfnamefont
  {D.}~\bibnamefont {Ziemkiewicz}},\ and\ \bibinfo {author} {\bibfnamefont
  {G.}~\bibnamefont {Czajkowski}},\ }\bibfield  {title} {\bibinfo {title}
  {Electro-optical properties of {{Rydberg}} excitons},\ }\href
  {https://doi.org/10.1103/PhysRevB.94.045205} {\bibfield  {journal} {\bibinfo
  {journal} {Physical Review B}\ }\textbf {\bibinfo {volume} {94}},\ \bibinfo
  {pages} {045205} (\bibinfo {year} {2016}{\natexlab{a}})}\BibitemShut
  {NoStop}%
\bibitem [{\citenamefont {{Zieli{\'n}ska-Raczy{\'n}ska}}\ \emph
  {et~al.}(2017)\citenamefont {{Zieli{\'n}ska-Raczy{\'n}ska}}, \citenamefont
  {Ziemkiewicz},\ and\ \citenamefont
  {Czajkowski}}]{zielinska-raczynskaMagnetoopticalPropertiesRydberg2017}%
  \BibitemOpen
  \bibfield  {author} {\bibinfo {author} {\bibfnamefont {S.}~\bibnamefont
  {{Zieli{\'n}ska-Raczy{\'n}ska}}}, \bibinfo {author} {\bibfnamefont
  {D.}~\bibnamefont {Ziemkiewicz}},\ and\ \bibinfo {author} {\bibfnamefont
  {G.}~\bibnamefont {Czajkowski}},\ }\bibfield  {title} {\bibinfo {title}
  {Magneto-optical properties of {{Rydberg}} excitons: {{Center-of-mass}}
  quantization approach},\ }\href {https://doi.org/10.1103/PhysRevB.95.075204}
  {\bibfield  {journal} {\bibinfo  {journal} {Physical Review B}\ }\textbf
  {\bibinfo {volume} {95}},\ \bibinfo {pages} {075204} (\bibinfo {year}
  {2017})}\BibitemShut {NoStop}%
\bibitem [{\citenamefont {Kurz}\ \emph {et~al.}(2017)\citenamefont {Kurz},
  \citenamefont {Grünwald},\ and\ \citenamefont {Scheel}}]{kurzexcitonic2017}%
  \BibitemOpen
  \bibfield  {author} {\bibinfo {author} {\bibfnamefont {M.}~\bibnamefont
  {Kurz}}, \bibinfo {author} {\bibfnamefont {P.}~\bibnamefont {Grünwald}},\
  and\ \bibinfo {author} {\bibfnamefont {S.}~\bibnamefont {Scheel}},\
  }\bibfield  {title} {\bibinfo {title} {Excitonic giant-dipole potentials in
  cuprous oxide},\ }\href {https://doi.org/10.1103/PhysRevB.95.245205}
  {\bibfield  {journal} {\bibinfo  {journal} {Physical Review B}\ }\textbf
  {\bibinfo {volume} {95}},\ \bibinfo {pages} {245205} (\bibinfo {year}
  {2017})}\BibitemShut {NoStop}%
\bibitem [{\citenamefont {A{\ss}mann}\ \emph {et~al.}(2016)\citenamefont
  {A{\ss}mann}, \citenamefont {Thewes}, \citenamefont {Fr{\"o}hlich},\ and\
  \citenamefont {Bayer}}]{assmannQuantumChaosBreaking2016}%
  \BibitemOpen
  \bibfield  {author} {\bibinfo {author} {\bibfnamefont {M.}~\bibnamefont
  {A{\ss}mann}}, \bibinfo {author} {\bibfnamefont {J.}~\bibnamefont {Thewes}},
  \bibinfo {author} {\bibfnamefont {D.}~\bibnamefont {Fr{\"o}hlich}},\ and\
  \bibinfo {author} {\bibfnamefont {M.}~\bibnamefont {Bayer}},\ }\bibfield
  {title} {\bibinfo {title} {Quantum chaos and breaking of all anti-unitary
  symmetries in {{Rydberg}} excitons},\ }\href
  {https://doi.org/10.1038/nmat4622} {\bibfield  {journal} {\bibinfo  {journal}
  {Nature Materials}\ }\textbf {\bibinfo {volume} {15}},\ \bibinfo {pages}
  {741} (\bibinfo {year} {2016})}\BibitemShut {NoStop}%
\bibitem [{\citenamefont {Schweiner}\ \emph
  {et~al.}(2017{\natexlab{b}})\citenamefont {Schweiner}, \citenamefont {Main},\
  and\ \citenamefont {Wunner}}]{schweinerMagnetoexcitonsBreakAntiunitary2017}%
  \BibitemOpen
  \bibfield  {author} {\bibinfo {author} {\bibfnamefont {F.}~\bibnamefont
  {Schweiner}}, \bibinfo {author} {\bibfnamefont {J.}~\bibnamefont {Main}},\
  and\ \bibinfo {author} {\bibfnamefont {G.}~\bibnamefont {Wunner}},\
  }\bibfield  {title} {\bibinfo {title} {Magnetoexcitons {{Break Antiunitary
  Symmetries}}},\ }\href {https://doi.org/10.1103/PhysRevLett.118.046401}
  {\bibfield  {journal} {\bibinfo  {journal} {Physical Review Letters}\
  }\textbf {\bibinfo {volume} {118}},\ \bibinfo {pages} {046401} (\bibinfo
  {year} {2017}{\natexlab{b}})}\BibitemShut {NoStop}%
\bibitem [{\citenamefont {Walther}\ \emph
  {et~al.}(2018{\natexlab{a}})\citenamefont {Walther}, \citenamefont {Krüger},
  \citenamefont {Scheel},\ and\ \citenamefont
  {Pohl}}]{waltherInteractionsRydbergExcitons2018a}%
  \BibitemOpen
  \bibfield  {author} {\bibinfo {author} {\bibfnamefont {V.}~\bibnamefont
  {Walther}}, \bibinfo {author} {\bibfnamefont {S.~O.}\ \bibnamefont
  {Krüger}}, \bibinfo {author} {\bibfnamefont {S.}~\bibnamefont {Scheel}},\
  and\ \bibinfo {author} {\bibfnamefont {T.}~\bibnamefont {Pohl}},\ }\bibfield
  {title} {\bibinfo {title} {Interactions between {Rydberg} excitons in
  {Cu}$_2${O}},\ }\href {https://doi.org/10.1103/PhysRevB.98.165201} {\bibfield
   {journal} {\bibinfo  {journal} {Physical Review B}\ }\textbf {\bibinfo
  {volume} {98}},\ \bibinfo {pages} {165201} (\bibinfo {year}
  {2018}{\natexlab{a}})}\BibitemShut {NoStop}%
\bibitem [{\citenamefont {Heck{\"o}tter}\ \emph {et~al.}(2021)\citenamefont
  {Heck{\"o}tter}, \citenamefont {Walther}, \citenamefont {Scheel},
  \citenamefont {Bayer}, \citenamefont {Pohl},\ and\ \citenamefont
  {A{\ss}mann}}]{heckotterAsymmetricRydbergBlockade2021}%
  \BibitemOpen
  \bibfield  {author} {\bibinfo {author} {\bibfnamefont {J.}~\bibnamefont
  {Heck{\"o}tter}}, \bibinfo {author} {\bibfnamefont {V.}~\bibnamefont
  {Walther}}, \bibinfo {author} {\bibfnamefont {S.}~\bibnamefont {Scheel}},
  \bibinfo {author} {\bibfnamefont {M.}~\bibnamefont {Bayer}}, \bibinfo
  {author} {\bibfnamefont {T.}~\bibnamefont {Pohl}},\ and\ \bibinfo {author}
  {\bibfnamefont {M.}~\bibnamefont {A{\ss}mann}},\ }\bibfield  {title}
  {\bibinfo {title} {Asymmetric {{Rydberg}} blockade of giant excitons in
  {{Cuprous Oxide}}},\ }\href {https://doi.org/10.1038/s41467-021-23852-z}
  {\bibfield  {journal} {\bibinfo  {journal} {Nature Communications}\ }\textbf
  {\bibinfo {volume} {12}},\ \bibinfo {pages} {3556} (\bibinfo {year}
  {2021})}\BibitemShut {NoStop}%
\bibitem [{\citenamefont {Semkat}\ \emph {et~al.}(2019)\citenamefont {Semkat},
  \citenamefont {Fehske},\ and\ \citenamefont
  {Stolz}}]{semkatInfluenceElectronholePlasma2019}%
  \BibitemOpen
  \bibfield  {author} {\bibinfo {author} {\bibfnamefont {D.}~\bibnamefont
  {Semkat}}, \bibinfo {author} {\bibfnamefont {H.}~\bibnamefont {Fehske}},\
  and\ \bibinfo {author} {\bibfnamefont {H.}~\bibnamefont {Stolz}},\ }\bibfield
   {title} {\bibinfo {title} {Influence of electron-hole plasma on {{Rydberg}}
  excitons in cuprous oxide},\ }\href
  {https://doi.org/10.1103/PhysRevB.100.155204} {\bibfield  {journal} {\bibinfo
   {journal} {Physical Review B}\ }\textbf {\bibinfo {volume} {100}},\ \bibinfo
  {pages} {155204} (\bibinfo {year} {2019})}\BibitemShut {NoStop}%
\bibitem [{\citenamefont {Semkat}\ \emph {et~al.}(2021)\citenamefont {Semkat},
  \citenamefont {Fehske},\ and\ \citenamefont
  {Stolz}}]{semkatQuantumManybodyEffects2021}%
  \BibitemOpen
  \bibfield  {author} {\bibinfo {author} {\bibfnamefont {D.}~\bibnamefont
  {Semkat}}, \bibinfo {author} {\bibfnamefont {H.}~\bibnamefont {Fehske}},\
  and\ \bibinfo {author} {\bibfnamefont {H.}~\bibnamefont {Stolz}},\ }\bibfield
   {title} {\bibinfo {title} {Quantum many-body effects on {{Rydberg}} excitons
  in cuprous oxide},\ }\href {https://doi.org/10.1140/epjs/s11734-021-00062-8}
  {\bibfield  {journal} {\bibinfo  {journal} {The European Physical Journal
  Special Topics}\ }\textbf {\bibinfo {volume} {230}},\ \bibinfo {pages} {947}
  (\bibinfo {year} {2021})}\BibitemShut {NoStop}%
\bibitem [{\citenamefont {Walther}\ and\ \citenamefont
  {Pohl}(2020)}]{waltherPlasmaEnhancedInteractionOptical2020}%
  \BibitemOpen
  \bibfield  {author} {\bibinfo {author} {\bibfnamefont {V.}~\bibnamefont
  {Walther}}\ and\ \bibinfo {author} {\bibfnamefont {T.}~\bibnamefont {Pohl}},\
  }\bibfield  {title} {\bibinfo {title} {Plasma-{{Enhanced Interaction}} and
  {{Optical Nonlinearities}} of {Cu}$_2${O} {{Rydberg Excitons}}},\ }\href
  {https://doi.org/10.1103/PhysRevLett.125.097401} {\bibfield  {journal}
  {\bibinfo  {journal} {Physical Review Letters}\ }\textbf {\bibinfo {volume}
  {125}},\ \bibinfo {pages} {097401} (\bibinfo {year} {2020})}\BibitemShut
  {NoStop}%
\bibitem [{\citenamefont {Morin}\ \emph {et~al.}(2022)\citenamefont {Morin},
  \citenamefont {Tignon}, \citenamefont {Mangeney}, \citenamefont {Dhillon},
  \citenamefont {Czajkowski}, \citenamefont {Karpi{\'n}ski}, \citenamefont
  {Zieli{\'n}ska-Raczy{\'n}ska}, \citenamefont {Ziemkiewicz},\ and\
  \citenamefont {Boulier}}]{morin_self-kerr_2022}%
  \BibitemOpen
  \bibfield  {author} {\bibinfo {author} {\bibfnamefont {C.}~\bibnamefont
  {Morin}}, \bibinfo {author} {\bibfnamefont {J.}~\bibnamefont {Tignon}},
  \bibinfo {author} {\bibfnamefont {J.}~\bibnamefont {Mangeney}}, \bibinfo
  {author} {\bibfnamefont {S.}~\bibnamefont {Dhillon}}, \bibinfo {author}
  {\bibfnamefont {G.}~\bibnamefont {Czajkowski}}, \bibinfo {author}
  {\bibfnamefont {K.}~\bibnamefont {Karpi{\'n}ski}}, \bibinfo {author}
  {\bibfnamefont {S.}~\bibnamefont {Zieli{\'n}ska-Raczy{\'n}ska}}, \bibinfo
  {author} {\bibfnamefont {D.}~\bibnamefont {Ziemkiewicz}},\ and\ \bibinfo
  {author} {\bibfnamefont {T.}~\bibnamefont {Boulier}},\ }\bibfield  {title}
  {\bibinfo {title} {Self-kerr {Effect} across the {Yellow Rydberg Series} of
  {Excitons} in {Cu}$_2${O}},\ }\href
  {https://doi.org/10.1103/PhysRevLett.129.137401} {\bibfield  {journal}
  {\bibinfo  {journal} {Physical Review Letters}\ }\textbf {\bibinfo {volume}
  {129}},\ \bibinfo {pages} {137401} (\bibinfo {year} {2022})}\BibitemShut
  {NoStop}%
\bibitem [{\citenamefont {{Zieli{\'n}ska-Raczy{\'n}ska}}\ \emph
  {et~al.}(2019)\citenamefont {{Zieli{\'n}ska-Raczy{\'n}ska}}, \citenamefont
  {Czajkowski}, \citenamefont {Karpi{\'n}ski},\ and\ \citenamefont
  {Ziemkiewicz}}]{zielinska-raczynskaNonlinearOpticalProperties2019}%
  \BibitemOpen
  \bibfield  {author} {\bibinfo {author} {\bibfnamefont {S.}~\bibnamefont
  {{Zieli{\'n}ska-Raczy{\'n}ska}}}, \bibinfo {author} {\bibfnamefont
  {G.}~\bibnamefont {Czajkowski}}, \bibinfo {author} {\bibfnamefont
  {K.}~\bibnamefont {Karpi{\'n}ski}},\ and\ \bibinfo {author} {\bibfnamefont
  {D.}~\bibnamefont {Ziemkiewicz}},\ }\bibfield  {title} {\bibinfo {title}
  {Nonlinear optical properties and self-{{Kerr}} effect of {{Rydberg}}
  excitons},\ }\href {https://doi.org/10.1103/PhysRevB.99.245206} {\bibfield
  {journal} {\bibinfo  {journal} {Physical Review B}\ }\textbf {\bibinfo
  {volume} {99}},\ \bibinfo {pages} {245206} (\bibinfo {year}
  {2019})}\BibitemShut {NoStop}%
\bibitem [{\citenamefont {{Zieli{\'n}ska-Raczy{\'n}ska}}\ \emph
  {et~al.}(2016{\natexlab{b}})\citenamefont {{Zieli{\'n}ska-Raczy{\'n}ska}},
  \citenamefont {Ziemkiewicz},\ and\ \citenamefont
  {Czajkowski}}]{zielinska-raczynskaElectromagneticallyInducedTransparency2016}%
  \BibitemOpen
  \bibfield  {author} {\bibinfo {author} {\bibfnamefont {S.}~\bibnamefont
  {{Zieli{\'n}ska-Raczy{\'n}ska}}}, \bibinfo {author} {\bibfnamefont
  {D.}~\bibnamefont {Ziemkiewicz}},\ and\ \bibinfo {author} {\bibfnamefont
  {G.}~\bibnamefont {Czajkowski}},\ }\bibfield  {title} {\bibinfo {title}
  {Electromagnetically {{Induced Transparency}} and slow light in media with
  {{Rydberg Excitons}}},\ }\href@noop {} {\bibfield  {journal} {\bibinfo
  {journal} {arXiv:1612.09170 [cond-mat]}\ } (\bibinfo {year}
  {2016}{\natexlab{b}})},\ \Eprint {https://arxiv.org/abs/1612.09170}
  {arXiv:1612.09170} \BibitemShut {NoStop}%
\bibitem [{\citenamefont {Walther}\ \emph {et~al.}(2020)\citenamefont
  {Walther}, \citenamefont {Grünwald},\ and\ \citenamefont
  {Pohl}}]{waltherElectromagneticallyInducedTransparency2020}%
  \BibitemOpen
  \bibfield  {author} {\bibinfo {author} {\bibfnamefont {V.}~\bibnamefont
  {Walther}}, \bibinfo {author} {\bibfnamefont {P.}~\bibnamefont {Grünwald}},\
  and\ \bibinfo {author} {\bibfnamefont {T.}~\bibnamefont {Pohl}},\ }\bibfield
  {title} {\bibinfo {title} {{Controlling} {Exciton}-{Phonon} {Interactions}
  via {Electromagnetically} {Induced} {Transparency}},\ }\href
  {https://doi.org/10.1103/PhysRevLett.125.173601} {\bibfield  {journal}
  {\bibinfo  {journal} {Physical Review Letters}\ }\textbf {\bibinfo {volume}
  {125}},\ \bibinfo {pages} {173601} (\bibinfo {year} {2020})}\BibitemShut
  {NoStop}%
\bibitem [{\citenamefont {Walther}\ \emph
  {et~al.}(2018{\natexlab{b}})\citenamefont {Walther}, \citenamefont {Johne},\
  and\ \citenamefont {Pohl}}]{waltherGiantOpticalNonlinearities2018a}%
  \BibitemOpen
  \bibfield  {author} {\bibinfo {author} {\bibfnamefont {V.}~\bibnamefont
  {Walther}}, \bibinfo {author} {\bibfnamefont {R.}~\bibnamefont {Johne}},\
  and\ \bibinfo {author} {\bibfnamefont {T.}~\bibnamefont {Pohl}},\ }\bibfield
  {title} {\bibinfo {title} {Giant optical nonlinearities from {{Rydberg}}
  excitons in semiconductor microcavities},\ }\href
  {https://doi.org/10.1038/s41467-018-03742-7} {\bibfield  {journal} {\bibinfo
  {journal} {Nature Communications}\ }\textbf {\bibinfo {volume} {9}},\
  \bibinfo {pages} {1309} (\bibinfo {year} {2018}{\natexlab{b}})}\BibitemShut
  {NoStop}%
\bibitem [{\citenamefont {Orfanakis}\ \emph {et~al.}(2022)\citenamefont
  {Orfanakis}, \citenamefont {Rajendran}, \citenamefont {Walther},
  \citenamefont {Volz}, \citenamefont {Pohl},\ and\ \citenamefont
  {Ohadi}}]{orfanakisRydbergExcitonPolaritons2022}%
  \BibitemOpen
  \bibfield  {author} {\bibinfo {author} {\bibfnamefont {K.}~\bibnamefont
  {Orfanakis}}, \bibinfo {author} {\bibfnamefont {S.~K.}\ \bibnamefont
  {Rajendran}}, \bibinfo {author} {\bibfnamefont {V.}~\bibnamefont {Walther}},
  \bibinfo {author} {\bibfnamefont {T.}~\bibnamefont {Volz}}, \bibinfo {author}
  {\bibfnamefont {T.}~\bibnamefont {Pohl}},\ and\ \bibinfo {author}
  {\bibfnamefont {H.}~\bibnamefont {Ohadi}},\ }\bibfield  {title} {\bibinfo
  {title} {Rydberg exciton\textendash polaritons in a {{Cu$_2$O}}
  microcavity},\ }\href {https://doi.org/10.1038/s41563-022-01230-4} {\bibfield
   {journal} {\bibinfo  {journal} {Nature Materials}\ ,\ \bibinfo {pages} {1}}
  (\bibinfo {year} {2022})}\BibitemShut {NoStop}%
\bibitem [{\citenamefont {Heckötter}\ \emph {et~al.}(2020)\citenamefont
  {Heckötter}, \citenamefont {Janas}, \citenamefont {Schwartz}, \citenamefont
  {Aßmann},\ and\ \citenamefont
  {Bayer}}]{heckotterExperimentalLimitationExtending2020}%
  \BibitemOpen
  \bibfield  {author} {\bibinfo {author} {\bibfnamefont {J.}~\bibnamefont
  {Heckötter}}, \bibinfo {author} {\bibfnamefont {D.}~\bibnamefont {Janas}},
  \bibinfo {author} {\bibfnamefont {R.}~\bibnamefont {Schwartz}}, \bibinfo
  {author} {\bibfnamefont {M.}~\bibnamefont {Aßmann}},\ and\ \bibinfo {author}
  {\bibfnamefont {M.}~\bibnamefont {Bayer}},\ }\bibfield  {title} {\bibinfo
  {title} {Experimental limitation in extending the exciton series in
  {Cu}$_{2}${O} towards higher principal quantum numbers},\ }\href
  {https://doi.org/10.1103/PhysRevB.101.235207} {\bibfield  {journal} {\bibinfo
   {journal} {Physical Review B}\ }\textbf {\bibinfo {volume} {101}},\ \bibinfo
  {pages} {235207} (\bibinfo {year} {2020})}\BibitemShut {NoStop}%
\bibitem [{\citenamefont {Versteegh}\ \emph {et~al.}(2021)\citenamefont
  {Versteegh}, \citenamefont {Steinhauer}, \citenamefont {Bajo}, \citenamefont
  {Lettner}, \citenamefont {Soro}, \citenamefont {Romanova}, \citenamefont
  {Gyger}, \citenamefont {Schweickert}, \citenamefont {Mysyrowicz},\ and\
  \citenamefont {Zwiller}}]{versteeghGiantRydbergExcitons2021}%
  \BibitemOpen
  \bibfield  {author} {\bibinfo {author} {\bibfnamefont {M.~A.~M.}\
  \bibnamefont {Versteegh}}, \bibinfo {author} {\bibfnamefont {S.}~\bibnamefont
  {Steinhauer}}, \bibinfo {author} {\bibfnamefont {J.}~\bibnamefont {Bajo}},
  \bibinfo {author} {\bibfnamefont {T.}~\bibnamefont {Lettner}}, \bibinfo
  {author} {\bibfnamefont {A.}~\bibnamefont {Soro}}, \bibinfo {author}
  {\bibfnamefont {A.}~\bibnamefont {Romanova}}, \bibinfo {author}
  {\bibfnamefont {S.}~\bibnamefont {Gyger}}, \bibinfo {author} {\bibfnamefont
  {L.}~\bibnamefont {Schweickert}}, \bibinfo {author} {\bibfnamefont
  {A.}~\bibnamefont {Mysyrowicz}},\ and\ \bibinfo {author} {\bibfnamefont
  {V.}~\bibnamefont {Zwiller}},\ }\bibfield  {title} {\bibinfo {title} {Giant
  {Rydberg} excitons in {Cu}$_2${O} probed by photoluminescence excitation
  spectroscopy},\ }\href {https://doi.org/10.1103/PhysRevB.104.245206}
  {\bibfield  {journal} {\bibinfo  {journal} {Physical Review B}\ }\textbf
  {\bibinfo {volume} {104}},\ \bibinfo {pages} {245206} (\bibinfo {year}
  {2021})}\BibitemShut {NoStop}%
\bibitem [{\citenamefont {Lynch}\ \emph {et~al.}(2021)\citenamefont {Lynch},
  \citenamefont {Hodges}, \citenamefont {Mandal}, \citenamefont {Langbein},
  \citenamefont {Singh}, \citenamefont {Gallagher}, \citenamefont {Pritchett},
  \citenamefont {Pizzey}, \citenamefont {Rogers}, \citenamefont {Adams},\ and\
  \citenamefont {Jones}}]{lynchRydbergExcitonsSynthetic2021}%
  \BibitemOpen
  \bibfield  {author} {\bibinfo {author} {\bibfnamefont {S.~A.}\ \bibnamefont
  {Lynch}}, \bibinfo {author} {\bibfnamefont {C.}~\bibnamefont {Hodges}},
  \bibinfo {author} {\bibfnamefont {S.}~\bibnamefont {Mandal}}, \bibinfo
  {author} {\bibfnamefont {W.}~\bibnamefont {Langbein}}, \bibinfo {author}
  {\bibfnamefont {R.~P.}\ \bibnamefont {Singh}}, \bibinfo {author}
  {\bibfnamefont {L.~A.~P.}\ \bibnamefont {Gallagher}}, \bibinfo {author}
  {\bibfnamefont {J.~D.}\ \bibnamefont {Pritchett}}, \bibinfo {author}
  {\bibfnamefont {D.}~\bibnamefont {Pizzey}}, \bibinfo {author} {\bibfnamefont
  {J.~P.}\ \bibnamefont {Rogers}}, \bibinfo {author} {\bibfnamefont {C.~S.}\
  \bibnamefont {Adams}},\ and\ \bibinfo {author} {\bibfnamefont {M.~P.~A.}\
  \bibnamefont {Jones}},\ }\bibfield  {title} {\bibinfo {title} {Rydberg
  excitons in synthetic cuprous oxide {Cu}$_2${O}},\ }\href
  {https://doi.org/10.1103/PhysRevMaterials.5.084602} {\bibfield  {journal}
  {\bibinfo  {journal} {Physical Review Materials}\ }\textbf {\bibinfo {volume}
  {5}},\ \bibinfo {pages} {084602} (\bibinfo {year} {2021})}\BibitemShut
  {NoStop}%
\bibitem [{\citenamefont {Kr{\"u}ger}\ \emph {et~al.}(2020)\citenamefont
  {Kr{\"u}ger}, \citenamefont {Stolz},\ and\ \citenamefont
  {Scheel}}]{krugerInteractionChargedImpurities2020}%
  \BibitemOpen
  \bibfield  {author} {\bibinfo {author} {\bibfnamefont {S.~O.}\ \bibnamefont
  {Kr{\"u}ger}}, \bibinfo {author} {\bibfnamefont {H.}~\bibnamefont {Stolz}},\
  and\ \bibinfo {author} {\bibfnamefont {S.}~\bibnamefont {Scheel}},\
  }\bibfield  {title} {\bibinfo {title} {Interaction of charged impurities and
  {{Rydberg}} excitons in cuprous oxide},\ }\href
  {https://doi.org/10.1103/PhysRevB.101.235204} {\bibfield  {journal} {\bibinfo
   {journal} {Physical Review B}\ }\textbf {\bibinfo {volume} {101}},\ \bibinfo
  {pages} {235204} (\bibinfo {year} {2020})}\BibitemShut {NoStop}%
\bibitem [{\citenamefont {Heck{\"o}tter}\ \emph {et~al.}(2017)\citenamefont
  {Heck{\"o}tter}, \citenamefont {Freitag}, \citenamefont {Fr{\"o}hlich},
  \citenamefont {A{\ss}mann}, \citenamefont {Bayer}, \citenamefont {Semina},\
  and\ \citenamefont {Glazov}}]{heckotterScalingLawsRydberg2017a}%
  \BibitemOpen
  \bibfield  {author} {\bibinfo {author} {\bibfnamefont {J.}~\bibnamefont
  {Heck{\"o}tter}}, \bibinfo {author} {\bibfnamefont {M.}~\bibnamefont
  {Freitag}}, \bibinfo {author} {\bibfnamefont {D.}~\bibnamefont
  {Fr{\"o}hlich}}, \bibinfo {author} {\bibfnamefont {M.}~\bibnamefont
  {A{\ss}mann}}, \bibinfo {author} {\bibfnamefont {M.}~\bibnamefont {Bayer}},
  \bibinfo {author} {\bibfnamefont {M.~A.}\ \bibnamefont {Semina}},\ and\
  \bibinfo {author} {\bibfnamefont {M.~M.}\ \bibnamefont {Glazov}},\ }\bibfield
   {title} {\bibinfo {title} {Scaling laws of {{Rydberg}} excitons},\ }\href
  {https://doi.org/10.1103/PhysRevB.96.125142} {\bibfield  {journal} {\bibinfo
  {journal} {Physical Review B}\ }\textbf {\bibinfo {volume} {96}},\ \bibinfo
  {pages} {125142} (\bibinfo {year} {2017})}\BibitemShut {NoStop}%
\bibitem [{\citenamefont
  {Elliott}(1957)}]{elliottIntensityOpticalAbsorption1957}%
  \BibitemOpen
  \bibfield  {author} {\bibinfo {author} {\bibfnamefont {R.~J.}\ \bibnamefont
  {Elliott}},\ }\bibfield  {title} {\bibinfo {title} {Intensity of {{Optical
  Absorption}} by {{Excitons}}},\ }\href@noop {} {\bibfield  {journal}
  {\bibinfo  {journal} {Physical Review}\ }\textbf {\bibinfo {volume} {108}},\
  \bibinfo {pages} {1384} (\bibinfo {year} {1957})}\BibitemShut {NoStop}%
\bibitem [{\citenamefont {Stolz}\ \emph {et~al.}(2018)\citenamefont {Stolz},
  \citenamefont {Sch{\"o}ne},\ and\ \citenamefont
  {Semkat}}]{stolzInteractionRydbergExcitons2018}%
  \BibitemOpen
  \bibfield  {author} {\bibinfo {author} {\bibfnamefont {H.}~\bibnamefont
  {Stolz}}, \bibinfo {author} {\bibfnamefont {F.}~\bibnamefont {Sch{\"o}ne}},\
  and\ \bibinfo {author} {\bibfnamefont {D.}~\bibnamefont {Semkat}},\
  }\bibfield  {title} {\bibinfo {title} {Interaction of {Rydberg} excitons in
  cuprous oxide with phonons and photons: optical linewidth and polariton
  effect},\ }\href {https://doi.org/10.1088/1367-2630/aaa396} {\bibfield
  {journal} {\bibinfo  {journal} {New Journal of Physics}\ }\textbf {\bibinfo
  {volume} {20}},\ \bibinfo {pages} {023019} (\bibinfo {year}
  {2018})}\BibitemShut {NoStop}%
\bibitem [{\citenamefont {Reigue}\ \emph {et~al.}(2019)\citenamefont {Reigue},
  \citenamefont {Hostein},\ and\ \citenamefont
  {Voliotis}}]{reigueResonanceFluorescenceSingle2019}%
  \BibitemOpen
  \bibfield  {author} {\bibinfo {author} {\bibfnamefont {A.}~\bibnamefont
  {Reigue}}, \bibinfo {author} {\bibfnamefont {R.}~\bibnamefont {Hostein}},\
  and\ \bibinfo {author} {\bibfnamefont {V.}~\bibnamefont {Voliotis}},\
  }\bibfield  {title} {\bibinfo {title} {Resonance fluorescence of a single
  semiconductor quantum dot: The impact of a fluctuating electrostatic
  environment},\ }\href {https://doi.org/10.1088/1361-6641/ab4362} {\bibfield
  {journal} {\bibinfo  {journal} {Semiconductor Science and Technology}\
  }\textbf {\bibinfo {volume} {34}},\ \bibinfo {pages} {113001} (\bibinfo
  {year} {2019})}\BibitemShut {NoStop}%
\bibitem [{\citenamefont {Nguyen}\ \emph {et~al.}(2012)\citenamefont {Nguyen},
  \citenamefont {Sallen}, \citenamefont {Voisin}, \citenamefont {Roussignol},
  \citenamefont {Diederichs},\ and\ \citenamefont
  {Cassabois}}]{nguyenOpticallyGatedResonant2012}%
  \BibitemOpen
  \bibfield  {author} {\bibinfo {author} {\bibfnamefont {H.~S.}\ \bibnamefont
  {Nguyen}}, \bibinfo {author} {\bibfnamefont {G.}~\bibnamefont {Sallen}},
  \bibinfo {author} {\bibfnamefont {C.}~\bibnamefont {Voisin}}, \bibinfo
  {author} {\bibfnamefont {P.}~\bibnamefont {Roussignol}}, \bibinfo {author}
  {\bibfnamefont {C.}~\bibnamefont {Diederichs}},\ and\ \bibinfo {author}
  {\bibfnamefont {G.}~\bibnamefont {Cassabois}},\ }\bibfield  {title} {\bibinfo
  {title} {Optically {{Gated Resonant Emission}} of {{Single Quantum Dots}}},\
  }\href {https://doi.org/10.1103/PhysRevLett.108.057401} {\bibfield  {journal}
  {\bibinfo  {journal} {Physical Review Letters}\ }\textbf {\bibinfo {volume}
  {108}},\ \bibinfo {pages} {057401} (\bibinfo {year} {2012})}\BibitemShut
  {NoStop}%
\bibitem [{\citenamefont {Houel}\ \emph {et~al.}(2012)\citenamefont {Houel},
  \citenamefont {Kuhlmann}, \citenamefont {Greuter}, \citenamefont {Xue},
  \citenamefont {Poggio}, \citenamefont {Gerardot}, \citenamefont {Dalgarno},
  \citenamefont {Badolato}, \citenamefont {Petroff}, \citenamefont {Ludwig},
  \citenamefont {Reuter}, \citenamefont {Wieck},\ and\ \citenamefont
  {Warburton}}]{houelProbingSingleChargeFluctuations2012}%
  \BibitemOpen
  \bibfield  {author} {\bibinfo {author} {\bibfnamefont {J.}~\bibnamefont
  {Houel}}, \bibinfo {author} {\bibfnamefont {A.~V.}\ \bibnamefont {Kuhlmann}},
  \bibinfo {author} {\bibfnamefont {L.}~\bibnamefont {Greuter}}, \bibinfo
  {author} {\bibfnamefont {F.}~\bibnamefont {Xue}}, \bibinfo {author}
  {\bibfnamefont {M.}~\bibnamefont {Poggio}}, \bibinfo {author} {\bibfnamefont
  {B.~D.}\ \bibnamefont {Gerardot}}, \bibinfo {author} {\bibfnamefont {P.~A.}\
  \bibnamefont {Dalgarno}}, \bibinfo {author} {\bibfnamefont {A.}~\bibnamefont
  {Badolato}}, \bibinfo {author} {\bibfnamefont {P.~M.}\ \bibnamefont
  {Petroff}}, \bibinfo {author} {\bibfnamefont {A.}~\bibnamefont {Ludwig}},
  \bibinfo {author} {\bibfnamefont {D.}~\bibnamefont {Reuter}}, \bibinfo
  {author} {\bibfnamefont {A.~D.}\ \bibnamefont {Wieck}},\ and\ \bibinfo
  {author} {\bibfnamefont {R.~J.}\ \bibnamefont {Warburton}},\ }\bibfield
  {title} {\bibinfo {title} {Probing {{Single-Charge Fluctuations}} at a
  {{GaAs}}/{{AlAs}} {{Interface Using Laser Spectroscopy}} on a {{Nearby InGaAs
  Quantum Dot}}},\ }\href {https://doi.org/10.1103/PhysRevLett.108.107401}
  {\bibfield  {journal} {\bibinfo  {journal} {Physical Review Letters}\
  }\textbf {\bibinfo {volume} {108}},\ \bibinfo {pages} {107401} (\bibinfo
  {year} {2012})}\BibitemShut {NoStop}%
\bibitem [{\citenamefont {Hauck}\ \emph {et~al.}(2014)\citenamefont {Hauck},
  \citenamefont {Seilmeier}, \citenamefont {Beavan}, \citenamefont {Badolato},
  \citenamefont {Petroff},\ and\ \citenamefont {Högele}}]{haucklocating2014}%
  \BibitemOpen
  \bibfield  {author} {\bibinfo {author} {\bibfnamefont {M.}~\bibnamefont
  {Hauck}}, \bibinfo {author} {\bibfnamefont {F.}~\bibnamefont {Seilmeier}},
  \bibinfo {author} {\bibfnamefont {S.~E.}\ \bibnamefont {Beavan}}, \bibinfo
  {author} {\bibfnamefont {A.}~\bibnamefont {Badolato}}, \bibinfo {author}
  {\bibfnamefont {P.~M.}\ \bibnamefont {Petroff}},\ and\ \bibinfo {author}
  {\bibfnamefont {A.}~\bibnamefont {Högele}},\ }\bibfield  {title} {\bibinfo
  {title} {Locating environmental charge impurities with confluent laser
  spectroscopy of multiple quantum dots},\ }\href
  {https://doi.org/10.1103/PhysRevB.90.235306} {\bibfield  {journal} {\bibinfo
  {journal} {Physical Review B}\ }\textbf {\bibinfo {volume} {90}},\ \bibinfo
  {pages} {235306} (\bibinfo {year} {2014})}\BibitemShut {NoStop}%
\bibitem [{\citenamefont {Chen}\ \emph {et~al.}(2016)\citenamefont {Chen},
  \citenamefont {Lander}, \citenamefont {Krowpman}, \citenamefont {Solomon},\
  and\ \citenamefont {Flagg}}]{chenCharacterizationLocalCharge2016}%
  \BibitemOpen
  \bibfield  {author} {\bibinfo {author} {\bibfnamefont {D.}~\bibnamefont
  {Chen}}, \bibinfo {author} {\bibfnamefont {G.~R.}\ \bibnamefont {Lander}},
  \bibinfo {author} {\bibfnamefont {K.~S.}\ \bibnamefont {Krowpman}}, \bibinfo
  {author} {\bibfnamefont {G.~S.}\ \bibnamefont {Solomon}},\ and\ \bibinfo
  {author} {\bibfnamefont {E.~B.}\ \bibnamefont {Flagg}},\ }\bibfield  {title}
  {\bibinfo {title} {Characterization of the local charge environment of a
  single quantum dot via resonance fluorescence},\ }\href
  {https://doi.org/10.1103/PhysRevB.93.115307} {\bibfield  {journal} {\bibinfo
  {journal} {Physical Review B}\ }\textbf {\bibinfo {volume} {93}},\ \bibinfo
  {pages} {115307} (\bibinfo {year} {2016})}\BibitemShut {NoStop}%
\bibitem [{\citenamefont {Makhonin}\ \emph {et~al.}(2014)\citenamefont
  {Makhonin}, \citenamefont {Dixon}, \citenamefont {Coles}, \citenamefont
  {Royall}, \citenamefont {Luxmoore}, \citenamefont {Clarke}, \citenamefont
  {Hugues}, \citenamefont {Skolnick},\ and\ \citenamefont
  {Fox}}]{makhoninwaveguide2014}%
  \BibitemOpen
  \bibfield  {author} {\bibinfo {author} {\bibfnamefont {M.~N.}\ \bibnamefont
  {Makhonin}}, \bibinfo {author} {\bibfnamefont {J.~E.}\ \bibnamefont {Dixon}},
  \bibinfo {author} {\bibfnamefont {R.~J.}\ \bibnamefont {Coles}}, \bibinfo
  {author} {\bibfnamefont {B.}~\bibnamefont {Royall}}, \bibinfo {author}
  {\bibfnamefont {I.~J.}\ \bibnamefont {Luxmoore}}, \bibinfo {author}
  {\bibfnamefont {E.}~\bibnamefont {Clarke}}, \bibinfo {author} {\bibfnamefont
  {M.}~\bibnamefont {Hugues}}, \bibinfo {author} {\bibfnamefont {M.~S.}\
  \bibnamefont {Skolnick}},\ and\ \bibinfo {author} {\bibfnamefont {A.~M.}\
  \bibnamefont {Fox}},\ }\bibfield  {title} {\bibinfo {title} {Waveguide
  {Coupled} {Resonance} {Fluorescence} from {On}-{Chip} {Quantum Emitter}},\
  }\href {https://doi.org/10.1021/nl5032937} {\bibfield  {journal} {\bibinfo
  {journal} {Nano Letters}\ }\textbf {\bibinfo {volume} {14}},\ \bibinfo
  {pages} {6997} (\bibinfo {year} {2014})}\BibitemShut {NoStop}%
\bibitem [{\citenamefont {Heck{\"o}tter}\ \emph {et~al.}(2018)\citenamefont
  {Heck{\"o}tter}, \citenamefont {Freitag}, \citenamefont {Fr{\"o}hlich},
  \citenamefont {A{\ss}mann}, \citenamefont {Bayer}, \citenamefont
  {Gr{\"u}nwald}, \citenamefont {Sch{\"o}ne}, \citenamefont {Semkat},
  \citenamefont {Stolz},\ and\ \citenamefont
  {Scheel}}]{heckotterRydbergExcitonsPresence2018}%
  \BibitemOpen
  \bibfield  {author} {\bibinfo {author} {\bibfnamefont {J.}~\bibnamefont
  {Heck{\"o}tter}}, \bibinfo {author} {\bibfnamefont {M.}~\bibnamefont
  {Freitag}}, \bibinfo {author} {\bibfnamefont {D.}~\bibnamefont
  {Fr{\"o}hlich}}, \bibinfo {author} {\bibfnamefont {M.}~\bibnamefont
  {A{\ss}mann}}, \bibinfo {author} {\bibfnamefont {M.}~\bibnamefont {Bayer}},
  \bibinfo {author} {\bibfnamefont {P.}~\bibnamefont {Gr{\"u}nwald}}, \bibinfo
  {author} {\bibfnamefont {F.}~\bibnamefont {Sch{\"o}ne}}, \bibinfo {author}
  {\bibfnamefont {D.}~\bibnamefont {Semkat}}, \bibinfo {author} {\bibfnamefont
  {H.}~\bibnamefont {Stolz}},\ and\ \bibinfo {author} {\bibfnamefont
  {S.}~\bibnamefont {Scheel}},\ }\bibfield  {title} {\bibinfo {title} {Rydberg
  {{Excitons}} in the {Presence} of an {Ultralow}-{{Density}}
  {{Electron}}-{{Hole}} {Plasma}},\ }\href
  {https://doi.org/10.1103/PhysRevLett.121.097401} {\bibfield  {journal}
  {\bibinfo  {journal} {Physical Review Letters}\ }\textbf {\bibinfo {volume}
  {121}},\ \bibinfo {pages} {097401} (\bibinfo {year} {2018})}\BibitemShut
  {NoStop}%
\bibitem [{\citenamefont {Bergen}\ \emph {et~al.}()\citenamefont {Bergen},
  \citenamefont {Walther}, \citenamefont {Panda}, \citenamefont {Harati},
  \citenamefont {Siegeroth}, \citenamefont {Heck{\"o}tter},\ and\ \citenamefont
  {A{\ss}mann}}]{bergenTamingChargedDefects2023}%
  \BibitemOpen
  \bibfield  {author} {\bibinfo {author} {\bibfnamefont {M.}~\bibnamefont
  {Bergen}}, \bibinfo {author} {\bibfnamefont {V.}~\bibnamefont {Walther}},
  \bibinfo {author} {\bibfnamefont {B.}~\bibnamefont {Panda}}, \bibinfo
  {author} {\bibfnamefont {M.}~\bibnamefont {Harati}}, \bibinfo {author}
  {\bibfnamefont {S.}~\bibnamefont {Siegeroth}}, \bibinfo {author}
  {\bibfnamefont {J.}~\bibnamefont {Heck{\"o}tter}},\ and\ \bibinfo {author}
  {\bibfnamefont {M.}~\bibnamefont {A{\ss}mann}},\ }\bibfield  {title}
  {\bibinfo {title} {Taming charged defects: Large scale purification in
  semiconductors using {Rydberg} {Excitons}},\ }\href
  {https://doi.org/10.48550/arXiv.2310.11726} {\bibinfo  {journal}
  {arXiv:2310.11726 [cond-mat.mes-hall]}\ }\BibitemShut {NoStop}%
\bibitem [{\citenamefont {Loison}\ \emph {et~al.}(1980)\citenamefont {Loison},
  \citenamefont {Robino},\ and\ \citenamefont {Schwab}}]{loison_progress_1980}%
  \BibitemOpen
\bibfield  {journal} {  }\bibfield  {author} {\bibinfo {author} {\bibfnamefont
  {J.~L.}\ \bibnamefont {Loison}}, \bibinfo {author} {\bibfnamefont
  {M.}~\bibnamefont {Robino}},\ and\ \bibinfo {author} {\bibfnamefont
  {C.}~\bibnamefont {Schwab}},\ }\bibfield  {title} {\bibinfo {title} {Progress
  in melt growth of {Cu}$_2${O}},\ }\href
  {https://doi.org/10.1016/0022-0248(80)90143-8} {\bibfield  {journal}
  {\bibinfo  {journal} {Journal of Crystal Growth}\ }\textbf {\bibinfo {volume}
  {50}},\ \bibinfo {pages} {816} (\bibinfo {year} {1980})}\BibitemShut
  {NoStop}%
\bibitem [{\citenamefont {Bloch}\ \emph {et~al.}(2080)\citenamefont {Bloch},
  \citenamefont {Meyer},\ and\ \citenamefont {Schwab}}]{bloch_sample_1980}%
  \BibitemOpen
  \bibfield  {author} {\bibinfo {author} {\bibfnamefont {P.~D.}\ \bibnamefont
  {Bloch}}, \bibinfo {author} {\bibfnamefont {B.}~\bibnamefont {Meyer}},\ and\
  \bibinfo {author} {\bibfnamefont {C.}~\bibnamefont {Schwab}},\ }\bibfield
  {title} {\bibinfo {title} {Sample thickness dependence of the exciton
  polariton absorption coefficient in {Cu}$_2${O}},\ }\href
  {https://doi.org/10.1088/0022-3719/13/2/014} {\bibfield  {journal} {\bibinfo
  {journal} {Journal of Physics C: Solid State Physics}\ }\textbf {\bibinfo
  {volume} {13}},\ \bibinfo {pages} {267} (\bibinfo {year} {2080})}\BibitemShut
  {NoStop}%
\bibitem [{\citenamefont {Thewes}\ \emph {et~al.}(2015)\citenamefont {Thewes},
  \citenamefont {Heck{\"o}tter}, \citenamefont {Kazimierczuk}, \citenamefont
  {A{\ss}mann}, \citenamefont {Fr{\"o}hlich}, \citenamefont {Bayer},
  \citenamefont {Semina},\ and\ \citenamefont
  {Glazov}}]{thewesObservationHighAngular2015}%
  \BibitemOpen
  \bibfield  {author} {\bibinfo {author} {\bibfnamefont {J.}~\bibnamefont
  {Thewes}}, \bibinfo {author} {\bibfnamefont {J.}~\bibnamefont
  {Heck{\"o}tter}}, \bibinfo {author} {\bibfnamefont {T.}~\bibnamefont
  {Kazimierczuk}}, \bibinfo {author} {\bibfnamefont {M.}~\bibnamefont
  {A{\ss}mann}}, \bibinfo {author} {\bibfnamefont {D.}~\bibnamefont
  {Fr{\"o}hlich}}, \bibinfo {author} {\bibfnamefont {M.}~\bibnamefont {Bayer}},
  \bibinfo {author} {\bibfnamefont {M.~A.}\ \bibnamefont {Semina}},\ and\
  \bibinfo {author} {\bibfnamefont {M.~M.}\ \bibnamefont {Glazov}},\ }\bibfield
   {title} {\bibinfo {title} {Observation of {{High Angular Momentum Excitons}}
  in {{Cuprous Oxide}}},\ }\href
  {https://doi.org/10.1103/PhysRevLett.115.027402} {\bibfield  {journal}
  {\bibinfo  {journal} {Physical Review Letters}\ }\textbf {\bibinfo {volume}
  {115}},\ \bibinfo {pages} {027402} (\bibinfo {year} {2015})}\BibitemShut
  {NoStop}%
\bibitem [{\citenamefont {Gr{\"u}nwald}\ \emph {et~al.}(2016)\citenamefont
  {Gr{\"u}nwald}, \citenamefont {A{\ss}mann}, \citenamefont {Heck{\"o}tter},
  \citenamefont {Fr{\"o}hlich}, \citenamefont {Bayer}, \citenamefont {Stolz},\
  and\ \citenamefont {Scheel}}]{grunwaldSignaturesQuantumCoherences2016}%
  \BibitemOpen
  \bibfield  {author} {\bibinfo {author} {\bibfnamefont {P.}~\bibnamefont
  {Gr{\"u}nwald}}, \bibinfo {author} {\bibfnamefont {M.}~\bibnamefont
  {A{\ss}mann}}, \bibinfo {author} {\bibfnamefont {J.}~\bibnamefont
  {Heck{\"o}tter}}, \bibinfo {author} {\bibfnamefont {D.}~\bibnamefont
  {Fr{\"o}hlich}}, \bibinfo {author} {\bibfnamefont {M.}~\bibnamefont {Bayer}},
  \bibinfo {author} {\bibfnamefont {H.}~\bibnamefont {Stolz}},\ and\ \bibinfo
  {author} {\bibfnamefont {S.}~\bibnamefont {Scheel}},\ }\bibfield  {title}
  {\bibinfo {title} {Signatures of {{Quantum Coherences}} in {{Rydberg
  Excitons}}},\ }\href {https://doi.org/10.1103/PhysRevLett.117.133003}
  {\bibfield  {journal} {\bibinfo  {journal} {Physical Review Letters}\
  }\textbf {\bibinfo {volume} {117}},\ \bibinfo {pages} {133003} (\bibinfo
  {year} {2016})}\BibitemShut {NoStop}%
\bibitem [{\citenamefont
  {Toyozawa}(1964)}]{toyozawaInterbandEffectLattice1964}%
  \BibitemOpen
  \bibfield  {author} {\bibinfo {author} {\bibfnamefont {Y.}~\bibnamefont
  {Toyozawa}},\ }\bibfield  {title} {\bibinfo {title} {Interband effect of
  lattice vibrations in the exciton absorption spectra},\ }\href
  {https://doi.org/doi:10.1016/0022-3697(64)90162-3} {\bibfield  {journal}
  {\bibinfo  {journal} {Journal of Physics and Chemistry of Solids}\ }\textbf
  {\bibinfo {volume} {25}},\ \bibinfo {pages} {59} (\bibinfo {year}
  {1964})}\BibitemShut {NoStop}%
\bibitem [{\citenamefont {Ueno}(1969)}]{uenoContourAbsorptionLines1969}%
  \BibitemOpen
  \bibfield  {author} {\bibinfo {author} {\bibfnamefont {T.}~\bibnamefont
  {Ueno}},\ }\bibfield  {title} {\bibinfo {title} {On the {Contour} of the
  {Absorption} {Lines} in {Cu}$_2${O}},\ }\href
  {https://doi.org/http://dx.doi.org/10.1143/JPSJ.26.438} {\bibfield  {journal}
  {\bibinfo  {journal} {Journal of the Physical Society of Japan}\ }\textbf
  {\bibinfo {volume} {26}},\ \bibinfo {pages} {438} (\bibinfo {year}
  {1969})}\BibitemShut {NoStop}%
\bibitem [{\citenamefont {Gallagher}(1994)}]{gallagherRydbergAtoms1994}%
  \BibitemOpen
  \bibfield  {author} {\bibinfo {author} {\bibfnamefont {T.~F.}\ \bibnamefont
  {Gallagher}},\ }\href@noop {} {\emph {\bibinfo {title} {{Rydberg Atoms}}}}\
  (\bibinfo  {publisher} {{Cambridge University Press}},\ \bibinfo {address}
  {{Cambridge}},\ \bibinfo {year} {1994})\BibitemShut {NoStop}%
\bibitem [{\citenamefont {Stolz}\ \emph {et~al.}(2022)\citenamefont {Stolz},
  \citenamefont {Semkat}, \citenamefont {Schwartz}, \citenamefont
  {Heck{\"o}tter}, \citenamefont {A{\ss}mann}, \citenamefont {Kraeft},
  \citenamefont {Fehske},\ and\ \citenamefont
  {Bayer}}]{stolzScrutinizingDebyePlasma2022}%
  \BibitemOpen
  \bibfield  {author} {\bibinfo {author} {\bibfnamefont {H.}~\bibnamefont
  {Stolz}}, \bibinfo {author} {\bibfnamefont {D.}~\bibnamefont {Semkat}},
  \bibinfo {author} {\bibfnamefont {R.}~\bibnamefont {Schwartz}}, \bibinfo
  {author} {\bibfnamefont {J.}~\bibnamefont {Heck{\"o}tter}}, \bibinfo {author}
  {\bibfnamefont {M.}~\bibnamefont {A{\ss}mann}}, \bibinfo {author}
  {\bibfnamefont {W.-D.}\ \bibnamefont {Kraeft}}, \bibinfo {author}
  {\bibfnamefont {H.}~\bibnamefont {Fehske}},\ and\ \bibinfo {author}
  {\bibfnamefont {M.}~\bibnamefont {Bayer}},\ }\bibfield  {title} {\bibinfo
  {title} {Scrutinizing the {{Debye}} plasma model: {{Rydberg}} excitons
  unravel the properties of low-density plasmas in semiconductors},\ }\href
  {https://doi.org/10.1103/PhysRevB.105.075204} {\bibfield  {journal} {\bibinfo
   {journal} {Physical Review B}\ }\textbf {\bibinfo {volume} {105}},\ \bibinfo
  {pages} {075204} (\bibinfo {year} {2022})}\BibitemShut {NoStop}%
\bibitem [{\citenamefont {Schöne}\ \emph {et~al.}(2017)\citenamefont
  {Schöne}, \citenamefont {Stolz},\ and\ \citenamefont
  {Naka}}]{schonePhononassistedAbsorptionExcitons2017}%
  \BibitemOpen
  \bibfield  {author} {\bibinfo {author} {\bibfnamefont {F.}~\bibnamefont
  {Schöne}}, \bibinfo {author} {\bibfnamefont {H.}~\bibnamefont {Stolz}},\
  and\ \bibinfo {author} {\bibfnamefont {N.}~\bibnamefont {Naka}},\ }\bibfield
  {title} {\bibinfo {title} {Phonon-assisted absorption of excitons in
  {{Cu}}$_{2}${O}},\ }\href {https://doi.org/10.1103/PhysRevB.96.115207}
  {\bibfield  {journal} {\bibinfo  {journal} {Physical Review B}\ }\textbf
  {\bibinfo {volume} {96}},\ \bibinfo {pages} {115207} (\bibinfo {year}
  {2017})}\BibitemShut {NoStop}%
\bibitem [{\citenamefont {Stolz}\ \emph {et~al.}(2021)\citenamefont {Stolz},
  \citenamefont {Schwartz}, \citenamefont {Heck{\"o}tter}, \citenamefont
  {A{\ss}mann}, \citenamefont {Semkat}, \citenamefont {Kr{\"u}ger},\ and\
  \citenamefont {Bayer}}]{stolzCoherentTransferMatrix2021}%
  \BibitemOpen
  \bibfield  {author} {\bibinfo {author} {\bibfnamefont {H.}~\bibnamefont
  {Stolz}}, \bibinfo {author} {\bibfnamefont {R.}~\bibnamefont {Schwartz}},
  \bibinfo {author} {\bibfnamefont {J.}~\bibnamefont {Heck{\"o}tter}}, \bibinfo
  {author} {\bibfnamefont {M.}~\bibnamefont {A{\ss}mann}}, \bibinfo {author}
  {\bibfnamefont {D.}~\bibnamefont {Semkat}}, \bibinfo {author} {\bibfnamefont
  {S.~O.}\ \bibnamefont {Kr{\"u}ger}},\ and\ \bibinfo {author} {\bibfnamefont
  {M.}~\bibnamefont {Bayer}},\ }\bibfield  {title} {\bibinfo {title} {Coherent
  transfer matrix analysis of the transmission spectra of {{Rydberg}} excitons
  in cuprous oxide},\ }\href {https://doi.org/10.1103/PhysRevB.104.035206}
  {\bibfield  {journal} {\bibinfo  {journal} {Physical Review B}\ }\textbf
  {\bibinfo {volume} {104}},\ \bibinfo {pages} {035206} (\bibinfo {year}
  {2021})}\BibitemShut {NoStop}%
\bibitem [{\citenamefont {Koirala}\ \emph {et~al.}(2014)\citenamefont
  {Koirala}, \citenamefont {Takahata}, \citenamefont {Hazama}, \citenamefont
  {Naka},\ and\ \citenamefont
  {Tanaka}}]{koiralaRelaxationLocalizedExcitons2014}%
  \BibitemOpen
  \bibfield  {author} {\bibinfo {author} {\bibfnamefont {S.}~\bibnamefont
  {Koirala}}, \bibinfo {author} {\bibfnamefont {M.}~\bibnamefont {Takahata}},
  \bibinfo {author} {\bibfnamefont {Y.}~\bibnamefont {Hazama}}, \bibinfo
  {author} {\bibfnamefont {N.}~\bibnamefont {Naka}},\ and\ \bibinfo {author}
  {\bibfnamefont {K.}~\bibnamefont {Tanaka}},\ }\bibfield  {title} {\bibinfo
  {title} {Relaxation of localized excitons by phonon emission at oxygen
  vacancies in {Cu}$_2${O}},\ }\href
  {https://doi.org/10.1016/j.jlumin.2014.06.027} {\bibfield  {journal}
  {\bibinfo  {journal} {Journal of Luminescence}\ }\textbf {\bibinfo {volume}
  {155}},\ \bibinfo {pages} {65} (\bibinfo {year} {2014})}\BibitemShut
  {NoStop}%
\bibitem [{\citenamefont {Frazer}\ \emph {et~al.}(2017)\citenamefont {Frazer},
  \citenamefont {Chang}, \citenamefont {Schaller}, \citenamefont
  {Poeppelmeier},\ and\ \citenamefont
  {Ketterson}}]{frazerVacancyRelaxationCuprous2017}%
  \BibitemOpen
  \bibfield  {author} {\bibinfo {author} {\bibfnamefont {L.}~\bibnamefont
  {Frazer}}, \bibinfo {author} {\bibfnamefont {K.~B.}\ \bibnamefont {Chang}},
  \bibinfo {author} {\bibfnamefont {R.~D.}\ \bibnamefont {Schaller}}, \bibinfo
  {author} {\bibfnamefont {K.~R.}\ \bibnamefont {Poeppelmeier}},\ and\ \bibinfo
  {author} {\bibfnamefont {J.~B.}\ \bibnamefont {Ketterson}},\ }\bibfield
  {title} {\bibinfo {title} {Vacancy relaxation in cuprous oxide
  ({Cu}$_{2-x}${O}$_{1-y}$)},\ }\href
  {https://doi.org/10.1016/j.jlumin.2016.11.011} {\bibfield  {journal}
  {\bibinfo  {journal} {Journal of Luminescence}\ }\textbf {\bibinfo {volume}
  {183}},\ \bibinfo {pages} {281} (\bibinfo {year} {2017})}\BibitemShut
  {NoStop}%
\bibitem [{\citenamefont {Takahata}\ and\ \citenamefont
  {Naka}(2018)}]{takahataPhotoluminescencePropertiesEntire2018a}%
  \BibitemOpen
  \bibfield  {author} {\bibinfo {author} {\bibfnamefont {M.}~\bibnamefont
  {Takahata}}\ and\ \bibinfo {author} {\bibfnamefont {N.}~\bibnamefont
  {Naka}},\ }\bibfield  {title} {\bibinfo {title} {Photoluminescence properties
  of the entire excitonic series in $\mathrm{Cu}_{2}\mathrm{O}$},\ }\href
  {https://doi.org/10.1103/PhysRevB.98.195205} {\bibfield  {journal} {\bibinfo
  {journal} {Physical Review B}\ }\textbf {\bibinfo {volume} {98}},\ \bibinfo
  {pages} {195205} (\bibinfo {year} {2018})}\BibitemShut {NoStop}%
\bibitem [{\citenamefont {Jang}\ \emph {et~al.}(2006)\citenamefont {Jang},
  \citenamefont {Sun}, \citenamefont {Watkins},\ and\ \citenamefont
  {Ketterson}}]{jangBoundExcitonsMathrmCu2006}%
  \BibitemOpen
  \bibfield  {author} {\bibinfo {author} {\bibfnamefont {J.~I.}\ \bibnamefont
  {Jang}}, \bibinfo {author} {\bibfnamefont {Y.}~\bibnamefont {Sun}}, \bibinfo
  {author} {\bibfnamefont {B.}~\bibnamefont {Watkins}},\ and\ \bibinfo {author}
  {\bibfnamefont {J.~B.}\ \bibnamefont {Ketterson}},\ }\bibfield  {title}
  {\bibinfo {title} {Bound excitons in {{Cu}}$_{2}${O}: Efficient internal free
  exciton detector},\ }\href {https://doi.org/10.1103/PhysRevB.74.235204}
  {\bibfield  {journal} {\bibinfo  {journal} {Physical Review B}\ }\textbf
  {\bibinfo {volume} {74}},\ \bibinfo {pages} {235204} (\bibinfo {year}
  {2006})}\BibitemShut {NoStop}%
\bibitem [{\citenamefont {Ito}\ and\ \citenamefont
  {Masumi}(1997)}]{itoDetailedExaminationRelaxation1997}%
  \BibitemOpen
  \bibfield  {author} {\bibinfo {author} {\bibfnamefont {T.}~\bibnamefont
  {Ito}}\ and\ \bibinfo {author} {\bibfnamefont {T.}~\bibnamefont {Masumi}},\
  }\bibfield  {title} {\bibinfo {title} {Detailed {{Examination}} of
  {{Relaxation Processes}} of {{Excitons}} in {{Photoluminescence Spectra}} of
  {Cu}$_2${O}},\ }\href {https://doi.org/10.1143/JPSJ.66.2185} {\bibfield
  {journal} {\bibinfo  {journal} {Journal of the Physical Society of Japan}\
  }\textbf {\bibinfo {volume} {66}},\ \bibinfo {pages} {2185} (\bibinfo {year}
  {1997})}\BibitemShut {NoStop}%
\bibitem [{\citenamefont {Steinhauer}\ \emph {et~al.}(2020)\citenamefont
  {Steinhauer}, \citenamefont {Versteegh}, \citenamefont {Gyger}, \citenamefont
  {Elshaari}, \citenamefont {Kunert}, \citenamefont {Mysyrowicz},\ and\
  \citenamefont {Zwiller}}]{steinhauerRydbergExcitonsCu2O2020}%
  \BibitemOpen
  \bibfield  {author} {\bibinfo {author} {\bibfnamefont {S.}~\bibnamefont
  {Steinhauer}}, \bibinfo {author} {\bibfnamefont {M.~A.~M.}\ \bibnamefont
  {Versteegh}}, \bibinfo {author} {\bibfnamefont {S.}~\bibnamefont {Gyger}},
  \bibinfo {author} {\bibfnamefont {A.~W.}\ \bibnamefont {Elshaari}}, \bibinfo
  {author} {\bibfnamefont {B.}~\bibnamefont {Kunert}}, \bibinfo {author}
  {\bibfnamefont {A.}~\bibnamefont {Mysyrowicz}},\ and\ \bibinfo {author}
  {\bibfnamefont {V.}~\bibnamefont {Zwiller}},\ }\bibfield  {title} {\bibinfo
  {title} {Rydberg excitons in {Cu}$_2${O} microcrystals grown on a silicon
  platform},\ }\href {https://doi.org/10.1038/s43246-020-0013-6} {\bibfield
  {journal} {\bibinfo  {journal} {Communications Materials}\ }\textbf {\bibinfo
  {volume} {1}},\ \bibinfo {pages} {11} (\bibinfo {year} {2020})}\BibitemShut
  {NoStop}%
\bibitem [{\citenamefont {Mund}\ \emph {et~al.}(2018)\citenamefont {Mund},
  \citenamefont {Fröhlich}, \citenamefont {Yakovlev},\ and\ \citenamefont
  {Bayer}}]{mundHighresolutionSecondHarmonic2018a}%
  \BibitemOpen
  \bibfield  {author} {\bibinfo {author} {\bibfnamefont {J.}~\bibnamefont
  {Mund}}, \bibinfo {author} {\bibfnamefont {D.}~\bibnamefont {Fröhlich}},
  \bibinfo {author} {\bibfnamefont {D.~R.}\ \bibnamefont {Yakovlev}},\ and\
  \bibinfo {author} {\bibfnamefont {M.}~\bibnamefont {Bayer}},\ }\bibfield
  {title} {\bibinfo {title} {High-resolution second harmonic generation
  spectroscopy with femtosecond laser pulses on excitons in {Cu}$_{2}${O}},\
  }\href {https://doi.org/10.1103/PhysRevB.98.085203} {\bibfield  {journal}
  {\bibinfo  {journal} {Physical Review B}\ }\textbf {\bibinfo {volume} {98}},\
  \bibinfo {pages} {085203} (\bibinfo {year} {2018})}\BibitemShut {NoStop}%
\bibitem [{\citenamefont {Rogers}\ \emph {et~al.}(2022)\citenamefont {Rogers},
  \citenamefont {Gallagher}, \citenamefont {Pizzey}, \citenamefont {Pritchett},
  \citenamefont {Adams}, \citenamefont {Jones}, \citenamefont {Hodges},
  \citenamefont {Langbein},\ and\ \citenamefont
  {Lynch}}]{rogersHighResolutionNanosecond2021}%
  \BibitemOpen
  \bibfield  {author} {\bibinfo {author} {\bibfnamefont {J.~P.}\ \bibnamefont
  {Rogers}}, \bibinfo {author} {\bibfnamefont {L.~A.~P.}\ \bibnamefont
  {Gallagher}}, \bibinfo {author} {\bibfnamefont {D.}~\bibnamefont {Pizzey}},
  \bibinfo {author} {\bibfnamefont {J.~D.}\ \bibnamefont {Pritchett}}, \bibinfo
  {author} {\bibfnamefont {C.~S.}\ \bibnamefont {Adams}}, \bibinfo {author}
  {\bibfnamefont {M.~P.~A.}\ \bibnamefont {Jones}}, \bibinfo {author}
  {\bibfnamefont {C.}~\bibnamefont {Hodges}}, \bibinfo {author} {\bibfnamefont
  {W.}~\bibnamefont {Langbein}},\ and\ \bibinfo {author} {\bibfnamefont
  {S.~A.}\ \bibnamefont {Lynch}},\ }\bibfield  {title} {\bibinfo {title}
  {High-resolution nanosecond spectroscopy of even-parity {Rydberg} excitons in
  {Cu$_2$O}},\ }\href {https://doi.org/10.1103/PhysRevB.105.115206} {\bibfield
  {journal} {\bibinfo  {journal} {Physical Review B}\ }\textbf {\bibinfo
  {volume} {105}},\ \bibinfo {pages} {115206} (\bibinfo {year}
  {2022})}\BibitemShut {NoStop}%
\bibitem [{\citenamefont {Gallagher}\ \emph {et~al.}(2022)\citenamefont
  {Gallagher}, \citenamefont {Rogers}, \citenamefont {Pritchett}, \citenamefont
  {Mistry}, \citenamefont {Pizzey}, \citenamefont {Adams}, \citenamefont
  {Jones}, \citenamefont {Grünwald}, \citenamefont {Walther}, \citenamefont
  {Hodges}, \citenamefont {Langbein},\ and\ \citenamefont
  {Lynch}}]{gallagher_microwave-optical_2022}%
  \BibitemOpen
  \bibfield  {author} {\bibinfo {author} {\bibfnamefont {L.~A.~P.}\
  \bibnamefont {Gallagher}}, \bibinfo {author} {\bibfnamefont {J.~P.}\
  \bibnamefont {Rogers}}, \bibinfo {author} {\bibfnamefont {J.~D.}\
  \bibnamefont {Pritchett}}, \bibinfo {author} {\bibfnamefont {R.~A.}\
  \bibnamefont {Mistry}}, \bibinfo {author} {\bibfnamefont {D.}~\bibnamefont
  {Pizzey}}, \bibinfo {author} {\bibfnamefont {C.~S.}\ \bibnamefont {Adams}},
  \bibinfo {author} {\bibfnamefont {M.~P.~A.}\ \bibnamefont {Jones}}, \bibinfo
  {author} {\bibfnamefont {P.}~\bibnamefont {Grünwald}}, \bibinfo {author}
  {\bibfnamefont {V.}~\bibnamefont {Walther}}, \bibinfo {author} {\bibfnamefont
  {C.}~\bibnamefont {Hodges}}, \bibinfo {author} {\bibfnamefont
  {W.}~\bibnamefont {Langbein}},\ and\ \bibinfo {author} {\bibfnamefont
  {S.~A.}\ \bibnamefont {Lynch}},\ }\bibfield  {title} {\bibinfo {title}
  {Microwave-optical coupling via {Rydberg} excitons in cuprous oxide},\ }\href
  {https://doi.org/10.1103/PhysRevResearch.4.013031} {\bibfield  {journal}
  {\bibinfo  {journal} {Physical Review Research}\ }\textbf {\bibinfo {volume}
  {4}},\ \bibinfo {pages} {013031} (\bibinfo {year} {2022})}\BibitemShut
  {NoStop}%
\bibitem [{\citenamefont {Orfanakis}\ \emph {et~al.}(2021)\citenamefont
  {Orfanakis}, \citenamefont {Rajendran}, \citenamefont {Ohadi}, \citenamefont
  {{Zieli{\'n}ska-Raczy{\'n}ska}}, \citenamefont {Czajkowski}, \citenamefont
  {Karpi{\'n}ski},\ and\ \citenamefont
  {Ziemkiewicz}}]{orfanakisQuantumConfinedRydberg2021}%
  \BibitemOpen
  \bibfield  {author} {\bibinfo {author} {\bibfnamefont {K.}~\bibnamefont
  {Orfanakis}}, \bibinfo {author} {\bibfnamefont {S.~K.}\ \bibnamefont
  {Rajendran}}, \bibinfo {author} {\bibfnamefont {H.}~\bibnamefont {Ohadi}},
  \bibinfo {author} {\bibfnamefont {S.}~\bibnamefont
  {{Zieli{\'n}ska-Raczy{\'n}ska}}}, \bibinfo {author} {\bibfnamefont
  {G.}~\bibnamefont {Czajkowski}}, \bibinfo {author} {\bibfnamefont
  {K.}~\bibnamefont {Karpi{\'n}ski}},\ and\ \bibinfo {author} {\bibfnamefont
  {D.}~\bibnamefont {Ziemkiewicz}},\ }\bibfield  {title} {\bibinfo {title}
  {Quantum confined {{Rydberg}} excitons in {Cu}$_2${O} nanoparticles},\ }\href
  {https://doi.org/10.1103/PhysRevB.103.245426} {\bibfield  {journal} {\bibinfo
   {journal} {Physical Review B}\ }\textbf {\bibinfo {volume} {103}},\ \bibinfo
  {pages} {245426} (\bibinfo {year} {2021})}\BibitemShut {NoStop}%
\bibitem [{\citenamefont {Naka}\ \emph {et~al.}(2005)\citenamefont {Naka},
  \citenamefont {Hashimoto},\ and\ \citenamefont
  {Ishihara}}]{nakaThinFilmsSingleCrystal2005}%
  \BibitemOpen
  \bibfield  {author} {\bibinfo {author} {\bibfnamefont {N.}~\bibnamefont
  {Naka}}, \bibinfo {author} {\bibfnamefont {S.}~\bibnamefont {Hashimoto}},\
  and\ \bibinfo {author} {\bibfnamefont {T.}~\bibnamefont {Ishihara}},\
  }\bibfield  {title} {\bibinfo {title} {Thin {{Films}} of {{Single-Crystal
  Cuprous Oxide Grown}} from the {{Melt}}},\ }\href
  {https://doi.org/10.1143/JJAP.44.5096} {\bibfield  {journal} {\bibinfo
  {journal} {Japanese Journal of Applied Physics}\ }\textbf {\bibinfo {volume}
  {44}},\ \bibinfo {pages} {5096} (\bibinfo {year} {2005})}\BibitemShut
  {NoStop}%
\bibitem [{\citenamefont {Chernikov}\ \emph {et~al.}(2014)\citenamefont
  {Chernikov}, \citenamefont {Berkelbach}, \citenamefont {Hill}, \citenamefont
  {Rigosi}, \citenamefont {Li}, \citenamefont {Aslan}, \citenamefont
  {Reichman}, \citenamefont {Hybertsen},\ and\ \citenamefont
  {Heinz}}]{chernikov_exciton_2014}%
  \BibitemOpen
  \bibfield  {author} {\bibinfo {author} {\bibfnamefont {A.}~\bibnamefont
  {Chernikov}}, \bibinfo {author} {\bibfnamefont {T.~C.}\ \bibnamefont
  {Berkelbach}}, \bibinfo {author} {\bibfnamefont {H.~M.}\ \bibnamefont
  {Hill}}, \bibinfo {author} {\bibfnamefont {A.}~\bibnamefont {Rigosi}},
  \bibinfo {author} {\bibfnamefont {Y.}~\bibnamefont {Li}}, \bibinfo {author}
  {\bibfnamefont {O.~B.}\ \bibnamefont {Aslan}}, \bibinfo {author}
  {\bibfnamefont {D.~R.}\ \bibnamefont {Reichman}}, \bibinfo {author}
  {\bibfnamefont {M.~S.}\ \bibnamefont {Hybertsen}},\ and\ \bibinfo {author}
  {\bibfnamefont {T.~F.}\ \bibnamefont {Heinz}},\ }\bibfield  {title} {\bibinfo
  {title} {Exciton {Binding} {Energy} and {Nonhydrogenic} {Rydberg} {Series} in
  {Monolayer} {WS}$_2$},\ }\href
  {https://doi.org/10.1103/PhysRevLett.113.076802} {\bibfield  {journal}
  {\bibinfo  {journal} {Physical Review Letters}\ }\textbf {\bibinfo {volume}
  {113}},\ \bibinfo {pages} {076802} (\bibinfo {year} {2014})}\BibitemShut
  {NoStop}%
\bibitem [{\citenamefont {Hill}\ \emph {et~al.}(2015)\citenamefont {Hill},
  \citenamefont {Rigosi}, \citenamefont {Roquelet}, \citenamefont {Chernikov},
  \citenamefont {Berkelbach}, \citenamefont {Reichman}, \citenamefont
  {Hybertsen}, \citenamefont {Brus},\ and\ \citenamefont
  {Heinz}}]{hill_observation_2015}%
  \BibitemOpen
  \bibfield  {author} {\bibinfo {author} {\bibfnamefont {H.~M.}\ \bibnamefont
  {Hill}}, \bibinfo {author} {\bibfnamefont {A.~F.}\ \bibnamefont {Rigosi}},
  \bibinfo {author} {\bibfnamefont {C.}~\bibnamefont {Roquelet}}, \bibinfo
  {author} {\bibfnamefont {A.}~\bibnamefont {Chernikov}}, \bibinfo {author}
  {\bibfnamefont {T.~C.}\ \bibnamefont {Berkelbach}}, \bibinfo {author}
  {\bibfnamefont {D.~R.}\ \bibnamefont {Reichman}}, \bibinfo {author}
  {\bibfnamefont {M.~S.}\ \bibnamefont {Hybertsen}}, \bibinfo {author}
  {\bibfnamefont {L.~E.}\ \bibnamefont {Brus}},\ and\ \bibinfo {author}
  {\bibfnamefont {T.~F.}\ \bibnamefont {Heinz}},\ }\bibfield  {title} {\bibinfo
  {title} {Observation of {Excitonic} {Rydberg} {States} in {Monolayer}
  {MoS}$_2$ and {WS}$_2$ by {Photoluminescence} {Excitation} {Spectroscopy}},\
  }\href {https://doi.org/10.1021/nl504868p} {\bibfield  {journal} {\bibinfo
  {journal} {Nano Letters}\ }\textbf {\bibinfo {volume} {15}},\ \bibinfo
  {pages} {2992} (\bibinfo {year} {2015})}\BibitemShut {NoStop}%
\end{thebibliography}
%


\end{document}